\documentclass[prd,twocolumn,showpacs,superscriptaddress,nofootinbib,floatfix,showkeys,10pt]{revtex4-2}
\usepackage{bm}
\usepackage{times}
\usepackage{braket}
\usepackage{amsfonts,amssymb,stmaryrd,latexsym,amsmath}
\usepackage[usenames,dvipsnames]{color}
\usepackage{slashed}
\usepackage{hyperref}
\usepackage{subfigure}
\usepackage{graphicx}
\usepackage{slashed}
\usepackage{orcidlink}
\usepackage{multirow}
\usepackage{appendix}
\usepackage{dashrule}
\usepackage{float}
\allowdisplaybreaks[1]

\definecolor{nicered}{rgb}{0.7,0.1,0.1}
\definecolor{nicegreen}{rgb}{0.1,0.5,0.1}
\definecolor{emph}{rgb}{1,0,0}
\definecolor{doub}{rgb}{0.7,0.2,1.0}
\definecolor{navyblue}{RGB}{0, 110, 184}

\hypersetup{colorlinks,citecolor=nicegreen,linkcolor=nicered,urlcolor=navyblue}

\newcommand{\clabel}[2][]{#2}

\begin{document}


	\title{Doubly heavy tetraquark bound and resonant states} 
	\author{Wei-Lin Wu\,\orcidlink{0009-0009-3480-8810}}\email{wlwu@pku.edu.cn}
	\affiliation{School of Physics, Peking University, Beijing 100871, China}
	\author{Yao Ma\,\orcidlink{0000-0002-5868-1166}}\email{yaoma@pku.edu.cn}
	\affiliation{School of Physics and Center of High Energy Physics,
		Peking University, Beijing 100871, China}
	\author{Yan-Ke Chen\,\orcidlink{0000-0002-9984-163X}}\email{chenyanke@stu.pku.edu.cn}
	\affiliation{School of Physics, Peking University, Beijing 100871, China}
	\author{Lu Meng\,\orcidlink{0000-0001-9791-7138}}\email{lu.meng@rub.de}
	\affiliation{Institut f\"ur Theoretische Physik II, Ruhr-Universit\"at Bochum,  D-44780 Bochum, Germany }
	\author{Shi-Lin Zhu\,\orcidlink{0000-0002-4055-6906}}\email{zhusl@pku.edu.cn}
	\affiliation{School of Physics and Center of High Energy Physics,
		Peking University, Beijing 100871, China}
	
	\begin{abstract}
		We calculate the energy spectrum of the S-wave doubly heavy tetraquark systems, including the $ QQ^{(\prime)}\bar q\bar q$,  $QQ^{(\prime)}\bar s\bar q$, and $ QQ^{(\prime)}\bar s\bar s$ ($Q^{(\prime)}=b,c$ and $q=u,d$) systems within the constituent quark model. We use the complex scaling method to obtain bound states and resonant states simultaneously, and the Gaussian expansion method to solve the complex-scaled four-body Schr\"odinger equation. With a novel definition of the root-mean-square radii, we are able to distinguish between meson molecules and compact tetraquark states. The compact tetraquarks are further classified into three different types with distinct spatial configurations: compact even tetraquarks, compact diquark-antidiquark tetraquarks and compact diquark-centered tetraquarks. In the $ I(J^P)=0(1^+) $ $QQ\bar q\bar q$ system, there exists the $ D^*D $ molecular bound state with a binding energy of $ -14 $ MeV, which is the candidate for $ T_{cc}(3875)^+ $. The shallow $\bar B^*\bar B$ molecular bound state is the bottom analog of $T_{cc}(3875)^+$. Moreover, we identify two resonant states near the $D^*D^*$ and $\bar B^*\bar B^*$ thresholds. In the $ J^P=1^+ $ $bb\bar q\bar q\,(I=0)$ and $bb\bar s\bar q$ systems, we obtain deeply bound states with a compact diquark-centered tetraquark configuration and a dominant $\chi_{\bar 3_c\otimes 3_c}$ component, along with resonant states with similar configurations as their radial excitations. These states are the QCD analog of the helium atom.  We also obtain some other bound states and resonant states with ``QCD hydrogen molecule" configurations. Moreover, we investigate the heavy quark mass dependence of the $ I(J^P)=0(1^+) $ $ QQ\bar q\bar q $ bound states. We strongly urge the experimental search for the predicted states.

	\end{abstract}
	
	\maketitle
	
	\section{Introduction}~\label{sec:intro}
	
	In 2021, the LHCb Collaboration discovered the first doubly charmed tetraquark state $ T_{cc}(3875)^+ $ in the $ D^0D^0\pi^+ $ invariant mass spectrum~\cite{LHCb:2021vvq,LHCb:2021auc}. It is a narrow state with a mass extremely close to the  $ D^{*+}D^0 $ threshold, having a binding energy of only around $ -300 $ keV. The observation of $ T_{cc}(3875)^+ $ significantly advances the hadron spectroscopy and may open a new chapter for the discovery of other doubly heavy exotic states in the future. 
	
	Theoretical investigations on the possible existence of doubly heavy tetraquark bound states dated back to the 1980s~\cite{Ader:1981db,Ballot:1983iv,Zouzou:1986qh,Carlson:1987hh}. Many studies aimed at predicting the masses of doubly charmed tetraquark states using various frameworks~\cite{Silvestre-Brac:1993zem,Pepin:1996id,Gelman:2002wf,Vijande:2003ki,Janc:2004qn,Cui:2006mp,Navarra:2007yw,Ebert:2007rn,Lee:2009rt,Yang:2009zzp,Dias:2011mi,Ohkoda:2012hv,Du:2012wp,Li:2012ss,Ikeda:2013vwa,Karliner:2017qjm,Eichten:2017ffp,Luo:2017eub,Wang:2017uld,Xu:2017tsr,Cheung:2017tnt,Park:2018wjk,Junnarkar:2018twb,Deng:2018kly,Liu:2019stu,Maiani:2019lpu,Yang:2019itm,Tan:2020ldi,Lu:2020rog,Meng:2020knc}, but their conclusions were wildly inconsistent, with the predicted masses ranging from $ -300 $ MeV to $ +300 $ MeV relative to the $DD^*$ threshold. After the discovery of $ T_{cc}(3875)^+ $, its exotic properties have attracted much attention and reignited interest in doubly heavy tetraquark states~\cite{Meng:2021jnw,Agaev:2021vur,Dong:2021bvy,Feijoo:2021ppq,Chen:2021vhg,Weng:2021hje,Xin:2021wcr,Chen:2021cfl,Albaladejo:2021vln,Du:2021zzh,Deng:2021gnb,Chen:2021spf,He:2022rta,Deng:2022cld,Meinel:2022lzo,Cheng:2022qcm,Wang:2022jop,Lyu:2023xro,Du:2023hlu,Hudspith:2023loy,Dai:2023kwv,Aoki:2023nzp,Wang:2023ovj,Padmanath:2023rdu,Meng:2023for,Wang:2023iaz,Ma:2023int,Meng:2023bmz,Mutuk:2023oyz,Abolnikov:2024key,Meng:2024yhu,Whyte:2024ihh,Ortega:2024epk,Colquhoun:2024jzh}. 
	The proximity of $ T_{cc}(3875)^+ $ to the $ D^{*+}D^0 $ threshold favors its interpretation as a $ D^*D $ molecular state, while for other doubly heavy tetraquark systems, both the compact tetraquark picture and the hadronic molecular picture have been proposed. More discussions can be found in recent reviews~\cite{Hosaka:2016pey,Ali:2017jda,Liu:2019zoy,Mai:2022eur,Meng:2022ozq,Chen:2022asf}.
	
	The discovery of the doubly charmed tetraquark bound state implies the existence of other doubly heavy tetraquark states. While many studies focus on the existence and properties of bound states, some also explore possible resonant states. In Ref.~\cite{Yang:2019itm}, the authors employed the complex scaling method to study the $ cc\bar q\bar q$, $bb\bar q\bar q$, $bc\bar q\bar q\,(q=u,d) $ bound and resonant states in the chiral quark model. However, they predicted a deeply bound $ cc\bar q\bar q $ state with a binding energy of around $ -150 $ MeV, which contradicts with the experimental results. In Ref.~\cite{Albaladejo:2021vln}, the author used the heavy quark spin symmetry to predict a resonant pole $ T_{cc}' $ below the $ D^*D^* $ channel as a partner of $ T_{cc}(3875) $. Similar results were obtained in the constituent quark model~\cite{Meng:2024yhu} and lattice QCD~\cite{Whyte:2024ihh}.  In Ref.~\cite{Meng:2023for}, the authors adopted a constituent quark model including the one-pion exchange interaction to study the $ cc\bar q\bar q$ and $ bb\bar q\bar q $ systems using the real scaling method. They did not find any resonant states in the doubly charmed sector, but reported a $ bb\bar q\bar q $ resonant state. 
	
	In this work, we conduct a comprehensive study on the S-wave doubly heavy tetraquark systems, including the $ QQ^{(\prime)}\bar q\bar q$,  $QQ^{(\prime)}\bar s\bar q$, and $ QQ^{(\prime)}\bar s\bar s$ ($Q^{(\prime)}=b,c$ and $q=u,d$)  systems, within the constituent quark model. We utilize the complex scaling method~\cite{Aguilar1971,Balslev1971,Aoyama2006} to obtain possible bound states and resonant states simultaneously. We employ the Gaussian expansion method~\cite{HIYAMA2003223} to solve the four-body Schrödinger equation, which has been successfully used in our previous work on tetraquark bound states~\cite{Meng:2023jqk} and resonant states~\cite{Chen:2023syh,Wu:2024euj,Wu:2024hrv,Ma:2024vsi}. Moreover, we calculate the root-mean-square radii of the tetraquark states to analyze their spatial structures and distinguish between meson molecular states and compact tetraquark states. 
    We further classify the compact tetraquark states into three different types with distinct spatial configurations, unraveling the rich internal structures and different forming mechanisms of the tetraquark states.
	
	This paper is organized as follows. In Sec.~\ref{sec:theo_framwork}, we introduce the theoretical framework, including the constituent quark model, the complex scaling method and the wave function construction. In Sec.~\ref{sec:spatial}, we demonstrate different spatial structures of tetraquarks and how to distinguish them by calculating the root-mean-square radii. In Sec.~\ref{sec:result}, we present the numerical results and discuss the properties of doubly heavy tetraquark states. We summarize our findings in Sec.~\ref{sec:summary}.

	\section{Theoretical Framework}\label{sec:theo_framwork}
	\subsection{Hamiltonian}
	In a nonrelativistic quark potential model, the Hamiltonian of a tetraquark system in the center-of-mass frame reads
	\begin{equation}
		H=\sum_{i=1}^4 (m_i+\frac{p_i^2}{2 m_i})+\sum_{i<j=1}^4 V_{ij
		},
	\end{equation}
	where the last term represents the two-body interaction between the $ i $-th and $ j $-th (anti)quark. \clabel[3b]{Though multi-body interactions may exist in the tetraquark system, it was shown that three-body interactions in baryon states may not have a significant effect, or they may be partially incorporated by the two-body interactions~\cite{Ma:2022vqf}. Therefore, the multi-body interactions in the tetraquark system may have a much smaller contribution than the two-body interactions and are not considered in this work.} We adopt the AL1 potential~\cite{Semay:1994ht,SilvestreBrac1996}, which includes the one-gluon-exchange interaction and a linear quark confinement interaction,
	\begin{equation}
		\begin{aligned}
			V_{i j} =-\frac{3}{16} \boldsymbol\lambda_i \cdot \boldsymbol\lambda_j\left(-\frac{\kappa}{r_{i j}}+\lambda r_{i j}-\Lambda\right. \\
			\left.+\frac{8 \pi \kappa^{\prime}}{3 m_i m_j} \frac{\exp \left(-r_{i j}^2 / r_0^2\right)}{\pi^{3 / 2} r_0^3} \boldsymbol{S}_i \cdot \boldsymbol{S}_j\right),
		\end{aligned}
	\end{equation}
	where $\boldsymbol\lambda_i$ and $ \boldsymbol{S}_i $ are the $\mathrm{SU}(3)$ color Gell-Mann matrix and the spin operator acting on quark $ i $, respectively. \clabel[cons]{The constant term $\sum_{i<j=1}^4\frac{3}{16}\Lambda \boldsymbol\lambda_i \cdot \boldsymbol\lambda_j$ determines the absolute value of energy but does not affect the energy gaps between different states.} \clabel[para]{The parameters of the AL1 model are listed in Table~\ref{tab:paraAL1}.} They were determined by fitting the meson spectra across all flavor sectors ~\cite{SilvestreBrac1996}. No additional free parameters are introduced. The theoretical masses as well as the root-mean-square (rms) radii of the corresponding mesons are listed in Table~\ref{tab:mesons}. It can be seen that the theoretical results for the $ 1S $ mesons agree with the experimental values within tens of MeV. We also list possible experimental candidates for the $ 2S $ excited mesons, whose theoretical masses deviate from the experimental values by up to 100 MeV. However, these candidates are not yet well established as $ 2S $ mesons. Moreover, relativistic effects and coupled-channel effects may play a crucial role in understanding these excited mesons, which is beyond the scope of this work. Nonetheless, since we focus only on the tetraquark states below the $ M(1S)M'(2S) $ dimeson thresholds in this work, we expect the uncertainties to be within tens of MeV, similar to those of the $ 1S $ mesons.
    \begin{table*}[htbp]
    \centering
    \caption{The parameters in the AL1 quark potential model.}
    \label{tab:paraAL1}
    \begin{tabular*}{\hsize}{@{}@{\extracolsep{\fill}}ccccccccccc@{}}
        \hline\hline
        $ \kappa $ &$ \lambda { [\mathrm{GeV}^{2}]}$&$ \Lambda {\rm [GeV]} $&$ \kappa^\prime $&$ m_b {\rm [GeV]}$&$ m_c {\rm [GeV]}$&$ m_s {\rm [GeV]}$&$ m_q {\rm [GeV]}$& $r_0 {\rm [GeV^{-1}]}$ &$ A { [\mathrm{GeV}^{B-1}]}$&$ B $\\
        \hline
        0.5069&0.1653&0.8321&1.8609&5.227&1.836&0.577&0.315&$A\left(\frac{2m_im_j}{m_i+m_j}\right)^{-B}$&1.6553&0.2204\\
        \hline\hline
    \end{tabular*}
\end{table*}
    
	\begin{table}[htbp]
		\centering
		\caption{The masses (in  $\mathrm{MeV}$) and rms radii (in fm) of singly heavy mesons in the AL1 quark model, compared with the experimental results taken from Ref.~\cite{ParticleDataGroup:2024cfk}. The ``$\dagger$" indicates possible experimental candidates for the $ 2S $ excited mesons~\cite{Chen:2016spr,Chen:2022asf}. }
		\label{tab:mesons}
		\begin{tabular}{lccc|lccc}
			\hline\hline
			Mesons& $ m_{\rm Exp.} $&$ m_{\rm AL1} $ &$ r^{\rm rms}_{\rm AL1} $& Mesons& $ m_{\rm Exp.} $&$ m_{\rm AL1} $ &$ r^{\rm rms}_{\rm AL1} $ \\
			\hline
			$D$ & $ 1867 $ &$  1862 $  & $ 0.61 $ & $D(2S)$ & $ 2549^\dagger $ &$  2643 $  & $ 1.23 $ \\
			$D^*$ & $ 2009 $ &$  2016 $  &$  0.70  $ &  $D^*(2S)$   & $ 2627^\dagger $ &$  2715  $ & $ 1.27 $ \\ 
			$D_s$ & $1968$ & $1965$ & $0.50$& $D_s(2S)$ & $ 2591^\dagger $ &$2663$ & $1.03$\\ 
			$D_s^*$	  & $2112$ & $2103$ & $0.57$&
			$D_s^*(2S)$ & $ 2714^\dagger $ &$2721$ & $1.06$\\
			$B$ & $ 5279 $ & $ 5293 $  & $ 0.62 $   &$B(2S)$ & $ 5863^\dagger $ &$  6013 $  & $ 1.20 $ \\  
			$B^*$ & $ 5325 $ & $ 5350 $ & $ 0.66 $ & $B^*(2S)$ & $ 5971^\dagger $ & $  6041  $  & $ 1.21 $ \\
			
			$B_s$ 	& $5367$ & $5362$ & $0.49$& $B_s(2S)$ 	&$ \cdots $ & $5998$ & $0.98$\\
			$B_s^*$ 	  & $5415$ & $5419$ & $0.52$
			& $B_s^*(2S)$ &$ \cdots $& $6022$ & $0.99$\\
		\hline\hline
		\end{tabular}
	\end{table}

	\subsection{Complex scaling method}
	In contrast to bound states, the wave functions of resonant states are not square integrable and cannot be obtained by solving the eigenequation of the Hermitian Hamiltonian directly. The complex scaling method (CSM) enables us to obtain possible bound states and resonant states simultaneously. In the CSM~\cite{Aguilar1971,Balslev1971,Aoyama2006}, the coordinate $ \boldsymbol{r} $ and its conjugate momentum $ \boldsymbol{p} $ are transformed as
	\begin{equation}
		U(\theta) \boldsymbol{r}=\boldsymbol{r} e^{i \theta}, \quad U(\theta) \boldsymbol{p}=\boldsymbol{p} e^{-i \theta}.
	\end{equation}
	Under such a transformation, the Hamiltonian is analytically continued to the complex plane and no longer Hermitian,
	\begin{equation}
		H(\theta)=\sum_{i=1}^4 (m_i+\frac{p_i^2e^{-2i\theta}}{2 m_i})+\sum_{i<j=1}^4 V_{ij}(r_{ij}e^{i\theta}).
	\end{equation}
	By solving the eigenvalue equation of the complex-scaled Hamiltonian, we can obtain the eigenenergies of bound states, scattering states and resonant states simultaneously. Bound states are located on the negative real axis in the energy plane and remain unchanged as $ \theta $ varies. Scattering states align along rays starting from threshold energies with $ \operatorname{Arg}(E)=-2\theta $.  If the complex scaling angle $ \theta $ is chosen to be larger than the angle of the resonant state $ \theta_r=\frac{1}{2}\tan^{-1}(\Gamma_r/2M_r) $, where $ M_r $ and $ \Gamma_r $ represent its mass and width, the wave function of the resonant state becomes square integrable and can be obtained with eigenenergy $ E_r=M_r-i\Gamma_r/2 $.

	\subsection{Wave function}
	The basis functions of the tetraquark wave function are expressed as
	\begin{equation}
		\psi=\mathcal{A}(\phi\otimes\chi),
	\end{equation}
	where $ \mathcal{A} $ is the antisymmetric operator of identical particles, and $ \phi $ and $ \chi $ represent the spatial wave function and color-spin wave function, respectively. 
	
\begin{figure}[htbp]
		\centering
		\includegraphics[width=.9\linewidth]{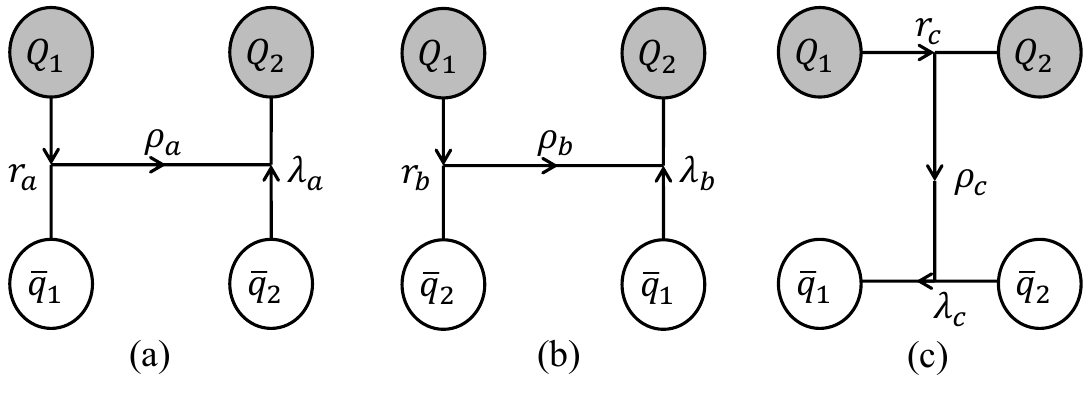}
		\caption{The Jacobian coordinates for two types of spatial configurations: (a), (b) for the dimeson configurations, and (c) for the diquark-antidiquark configuration.}
		\label{fig:jac}
	\end{figure}

	For the spatial wave function, we employ the Gaussian expansion method~\cite{HIYAMA2003223}. We consider three sets of spatial configurations (dimeson and diquark-antidiquark), which are denoted by $ (\rm{jac}) = (a),\,(b),\,(c) $. In each configuration, there are three independent Jacobian coordinates $r_{jac}, \lambda_{jac}$, $\rho_{jac}$, as shown in Fig.~\ref{fig:jac}. The S-wave spatial basis function is written as
	\begin{equation}
		\phi^{(\rm jac)}_{n_1,n_2,n_3}=\phi_{n_{1}}(r_{jac})\phi_{n_{2}}(\lambda_{jac})\phi_{n_{3}}(\rho_{jac}),
	\end{equation}
	where $ \phi_{n_i}(r) $ takes the Gaussian form,
	\begin{equation}\label{eq:Gaussian}
		\begin{aligned}
			&\phi_{n_i}(r)=N_{n_{i}}e^{-\nu_{n_i}r^2},\\
			&\nu_{n_i}=\nu_{1}\gamma^{n_i-1}\quad (n_i=1\sim n_{\rm max}),
		\end{aligned}
	\end{equation}
	and $ N_{n_i} $ is the normalization factor. \clabel[n]{We take $n_{\rm max}=12$ to obtain numerically stable results.}
	
	For the color-spin wave function, we choose a complete set of bases written as
	\begin{equation}\label{eq:colorspin_wf1}
		\begin{aligned}
			\chi^{s_1,s_2,S}_{\bar 3_c\otimes 3_c}=\left[\left(Q_1Q_2\right)_{\bar 3_c}^{s_1}\left(\bar q_1\bar q_2\right)_{3_c}^{s_2}\right]_{1_c}^{S},\\
			\chi^{s_1,s_2,S}_{6_c\otimes \bar 6_c}=\left[\left(Q_1Q_2\right)_{6_c}^{s_1}\left(\bar q_1\bar q_2\right)_{\bar{6}_c}^{s_2}\right]_{1_c}^{S},\\
		\end{aligned}
	\end{equation}
	where the subscripts and superscripts denote the color and spin representations, respectively.
	
	\section{Spatial structures}\label{sec:spatial}
\begin{figure*}
		\centering
		\includegraphics[width=0.9\linewidth]{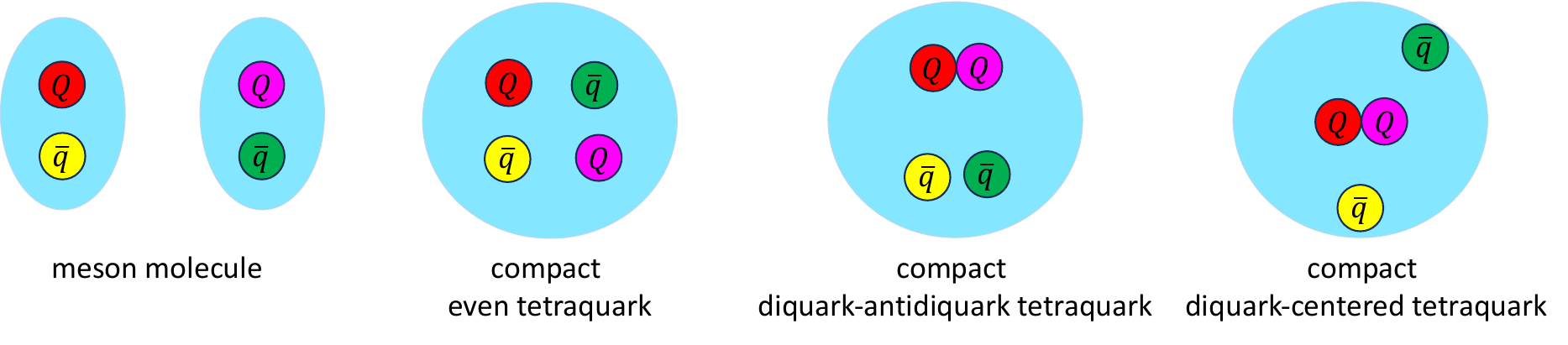}
		\caption{Categorization of spatial structures of tetraquark states. The blue background represents the range of the confinement. }
		\label{fig:config}
	\end{figure*}

	Tetraquark states are generally classified into meson molecules and compact tetraquarks. In the molecular scheme, the (anti)quarks cluster into two color-singlet mesons, and their relative distance is expected to be larger than the typical range of color confinement, $\Lambda_{QCD}^{-1}\sim 1$ fm, which is also the typical size of a meson. In contrast, in the compact tetraquark scheme, all four (anti)quarks are confined together, with their relative distances being on the order of $\Lambda_{QCD}^{-1}$. The compact tetraquark scheme can be further subdivided into several types, particularly for doubly heavy systems, as illustrated in Fig.~\ref{fig:config}. In compact even tetraquarks, the relative distance between four (anti)quarks is of similar size. In compact diquark-antidiquark tetraquarks, two quarks and two antiquarks form two clusters, respectively. The sizes of the clusters are smaller than the distance between them. In compact diquark-centered tetraquarks, the two heavy quarks form a very compact diquark cluster,  while the two light antiquarks orbit around the diquark, similar to the way two electrons orbit around the nucleus in a helium atom. The compact diquark-centered tetraquarks correspond to the ``QCD helium atom" in Ref.~\cite{Liu:2019zoy}. The compact even tetraquarks  roughly correspond the ``QCD hydrogen molecule" states as coined in Ref.~\cite{Liu:2019zoy}. The two light antiquarks are shared by two heavy quarks as in the hydrogen molecule where two electrons are shared by two protons.  It should be noted that in this paper (anti)diquark only refers to two (anti)quarks that form a spatially compact cluster, rather than the color-spin-isospin configuration of two (anti)quarks as discussed in Ref.~\cite{Jaffe:2004ph}.
	
	The rms radius is a commonly used criterion to analyze the spatial structures of tetraquark states and distinguish between different tetraquark configurations. However, we found that the conventional rms radii calculated using the complete wave function could be misleading and fail to identify molecular configuration due to the antisymmetrization of identical particles~\cite{Chen:2023syh,Wu:2024euj}. In order to eliminate the ambiguity, we proposed a new approach to calculate the rms radius. For systems without identical particle such as $ bc\bar s\bar q $, such an ambiguity does not exist and we can calculate the rms radii using the complete wave function directly. For systems with one pair of identical particles $\left( bc\bar q\bar q\,(qq\bar b\bar c) ,  bc\bar s\bar s\,(ss\bar b\bar c) , cc\bar s\bar q, bb\bar s\bar q \right)$, we decompose the complete antisymmetric wave function as 
	\begin{equation}\label{eq:wf_decompose1}
		\begin{aligned}
			\Psi(\theta)=&[(q_1\bar {q'})_{1_c}(q_2\bar{ q''})_{1_c}]_{1_c}\otimes|\psi_1(\theta)\rangle\\
			&+[(q_2\bar {q'})_{1_c}(q_1\bar{q''})_{1_c}]_{1_c}\otimes|\psi_2(\theta)\rangle\\
			=&\mathcal{A}\,[(q_1\bar {q'})_{1_c}(q_2\bar {q''})_{1_c}]_{1_c}\otimes|\psi_1(\theta)\rangle\\
			\equiv&\mathcal{A}\,\Psi_{\rm nA}(\theta).
		\end{aligned}
	\end{equation}
	For systems with two pairs of identical particles $ (cc\bar q\bar q, bb\bar q\bar q, cc\bar s\bar s, bb\bar s\bar s) $, we decompose the complete antisymmetric wave function as
	\begin{equation}\label{eq:wf_decompose2}
		\begin{aligned}
			\Psi(\theta)=&\sum_{s_1\geq s_2}\left([(Q_1\bar q_1)^{s_1}_{1_c}(Q_2\bar q_2)^{s_2}_{1_c}]^{S}_{1_c}\otimes|\psi_1^{s_1s_2}(\theta)\rangle\right.\\
			&+[(Q_1\bar q_1)^{s_2}_{1_c}(Q_2\bar q_2)^{s_1}_{1_c}]^{S}_{1_c}\otimes|\psi_2^{s_1s_2}(\theta)\rangle\\
			&+[(Q_1\bar q_2)^{s_1}_{1_c}(Q_2\bar q_1)^{s_2}_{1_c}]^{S}_{1_c}\otimes|\psi_3^{s_1s_2}(\theta)\rangle\\
			&+\left.[(Q_1\bar q_2)^{s_2}_{1_c}(Q_2\bar q_1)^{s_1}_{1_c}]^{S}_{1_c}\otimes|\psi_4^{s_1s_2}(\theta)\rangle\right)\\
			=&\mathcal{A}\sum_{s_1\geq s_2}[(Q_1\bar q_1)^{s_1}_{1_c}(Q_2\bar q_2)^{s_2}_{1_c}]^{S}_{1_c}\otimes|\psi_1^{s_1s_2}(\theta)\rangle\\
			\equiv&\mathcal{A}\,\Psi_{\rm nA}(\theta),
		\end{aligned}
	\end{equation}
	where $ s_1, s_2 $ sum over spin configurations with total spin $ S $. Instead of using the complete wave function $ \Psi(\theta) $, we use the decomposed nonantisymmetric wave function $ \Psi_{\mathrm{nA}}(\theta) $ to calculate the rms radius:
	\begin{equation}\label{eq:rmsr}
		r^{\mathrm{rms}}_{ij}\equiv \mathrm{Re}\left[\sqrt{\frac{\langle\Psi_{\mathrm{nA}}(\theta) | r_{ij}^2 e^{2i\theta}|\Psi_{\mathrm{nA}}(\theta)\rangle}{\langle\Psi_{\mathrm{nA}}(\theta) | \Psi_{\mathrm{nA}}(\theta)\rangle}}\right].
	\end{equation}
 For a meson molecule, $ r^{\mathrm{rms}}_{13} $ and $ r^{\mathrm{rms}}_{24} $ are expected to be the sizes of the constituent mesons, and much smaller than the other rms radii. The novel definition of the rms radius is useful for distinguishing the molecular configuration from the others.  For a compact tetraquark state, it may seem more reasonable to calculate the rms radii using the complete wave function. However, we find that the results from the conventional definition and the novel definition are qualitatively the same for compact tetraquarks, as shown in Appendix~\ref{app:rms}. Therefore, we will only use the novel definition to calculate the rms radii. 
	
The inner products in the CSM are defined using the c product~\cite{ROMO1968617}, 
	\begin{equation}
		\label{eq:cpro}
		\langle\phi_n \mid \phi_m\rangle\equiv\int \phi_n(r)\phi_{m}(r)d^3r,
	\end{equation}
where the complex conjugate of the ``bra" state is not taken. The c-product definition ensures that the expectation values of physical quantities are independent of the complex scaling angle  $ \theta $. The rms radius calculated by the c product is generally not real, but its real part can still reflect the internal quark clustering behavior if the width of the resonant state is not too large, as discussed in Refs.~\cite{Berggren:1970wto,homma1997matrix}.

\section{Results and Discussions}\label{sec:result}
\subsection{$ QQ^{(\prime)}\bar q\bar q $}
\begin{figure*}[htbp]
		\centering
		\includegraphics[width=1\linewidth]{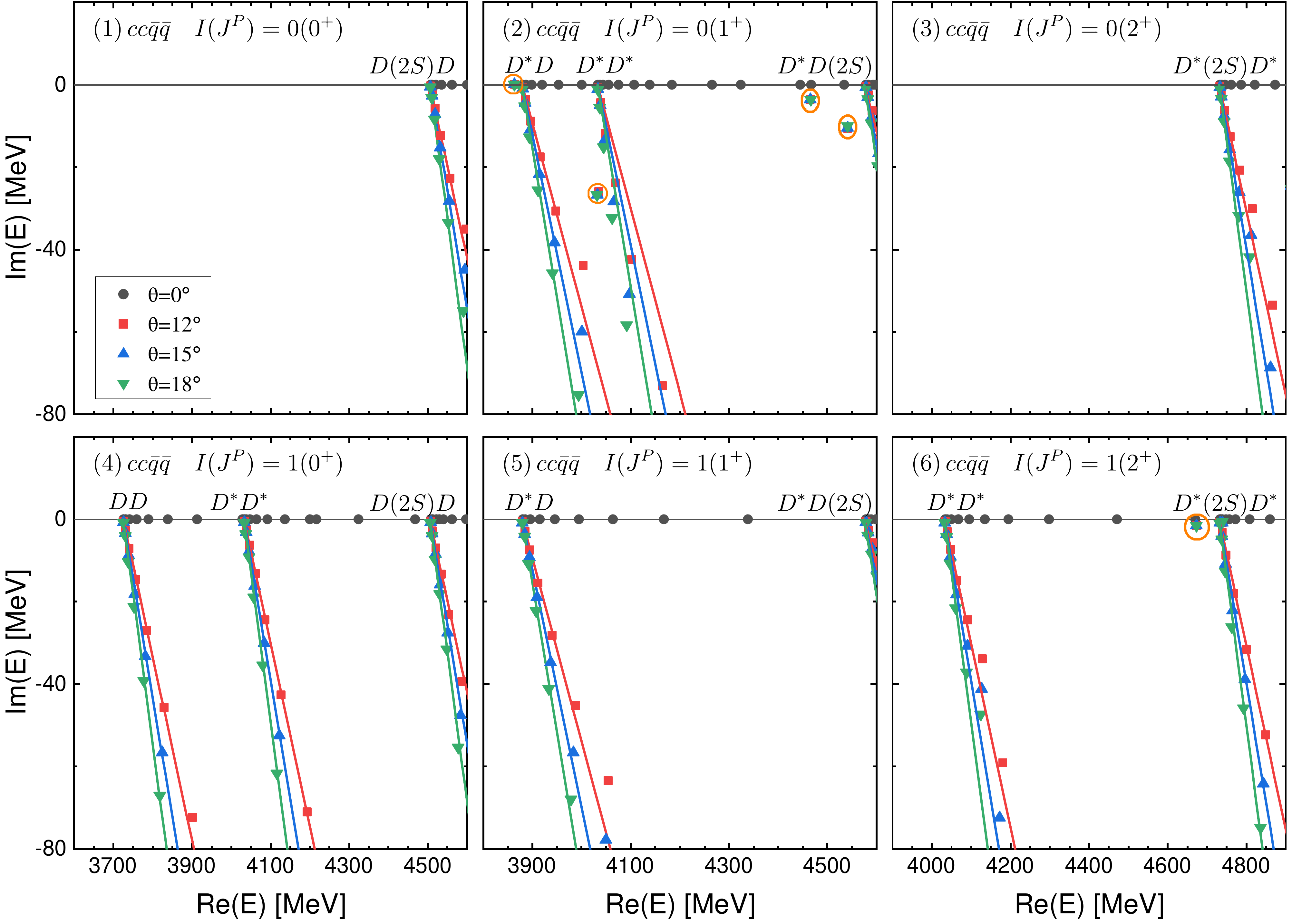}
		\caption{The complex energy eigenvalues of the  $cc\bar q\bar q$ states with varying $\theta$ in the CSM. The solid lines represent the continuum lines rotating along $\operatorname{Arg}(E)=-2 \theta$. The bound and resonant states do not shift as $\theta$ changes and are highlighted by the orange circles.}
		\label{fig:ccqq}
	\end{figure*}

\begin{figure*}[htbp]
		\centering
		\includegraphics[width=1\linewidth]{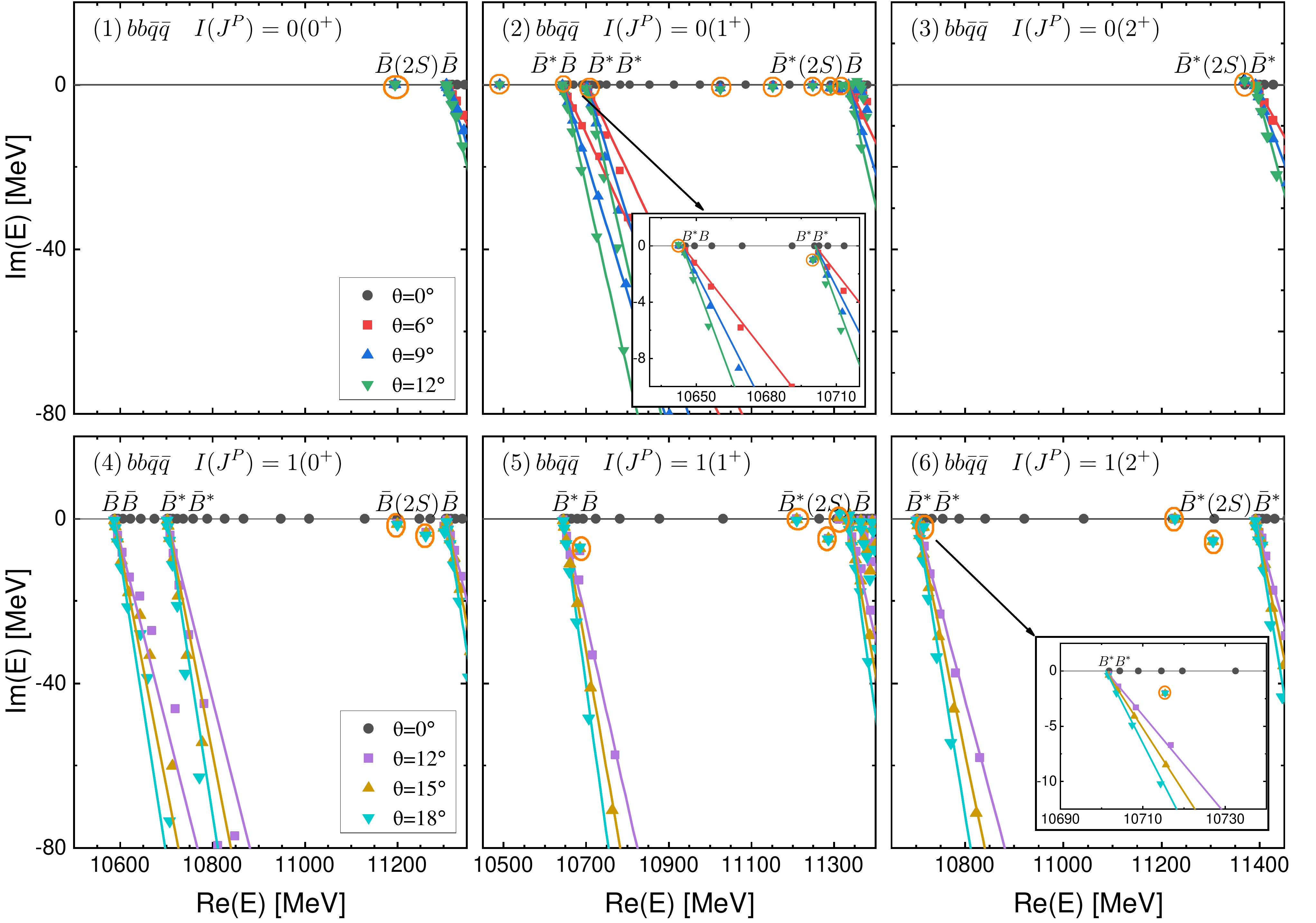}
		\caption{The complex energy eigenvalues of the  $bb\bar q\bar q$ states with varying $\theta$ in the CSM. The solid lines represent the continuum lines rotating along $\operatorname{Arg}(E)=-2 \theta$. The bound and resonant states do not shift as $\theta$ changes and are highlighted by the orange circles.}
		\label{fig:bbqq}
	\end{figure*}

\begin{figure*}[htbp]
		\centering
		\includegraphics[width=1\linewidth]{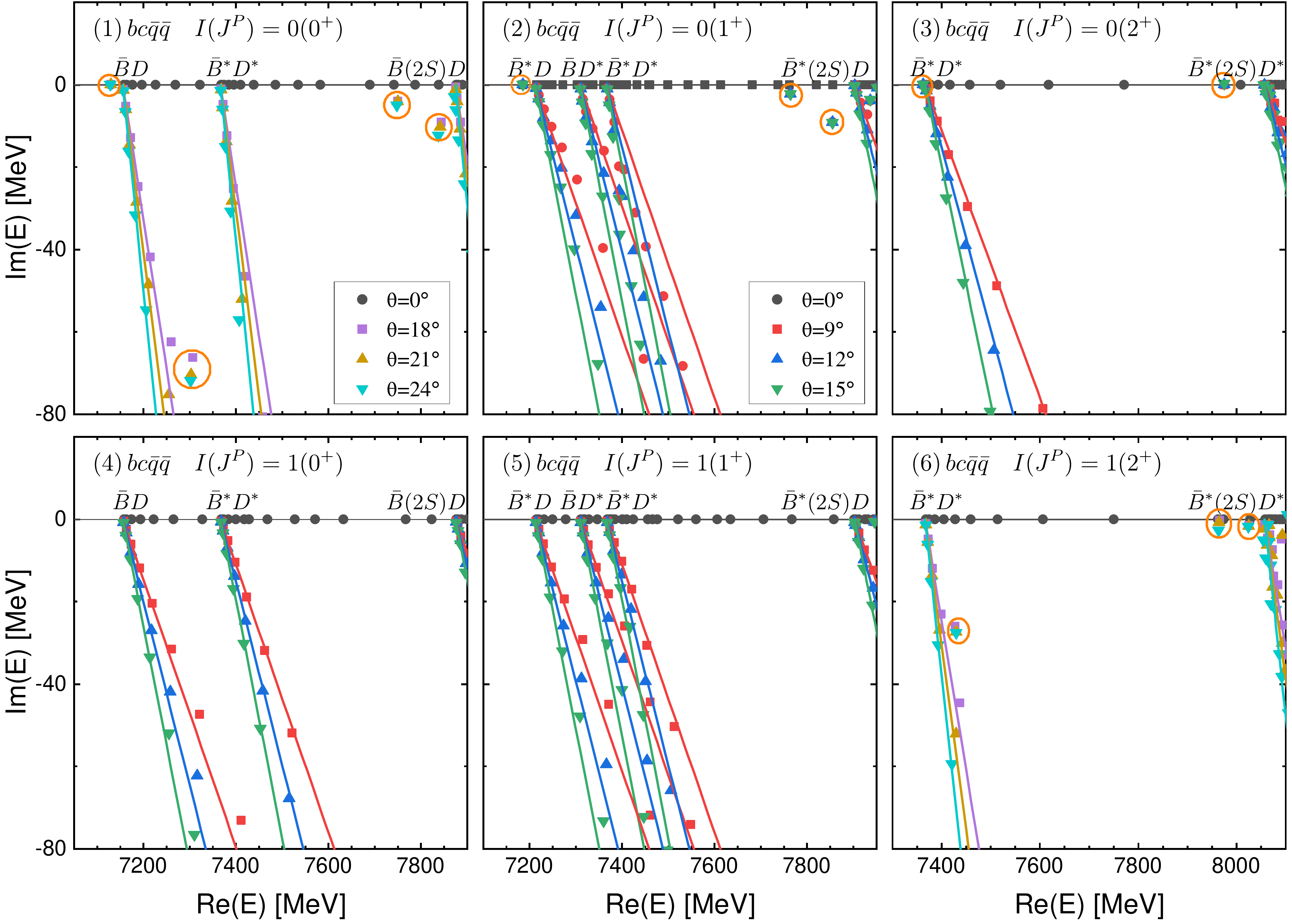}
		\caption{The complex energy eigenvalues of the  $bc\bar q\bar q$ states with varying $\theta$ in the CSM. The solid lines represent the continuum lines rotating along $\operatorname{Arg}(E)=-2 \theta$. The bound and resonant states do not shift as $\theta$ changes and are highlighted by the orange circles.}
		\label{fig:bcqq}
	\end{figure*}

With the CSM, we calculate the complex eigenenergies of the S-wave $ QQ^{(\prime)}\bar q\bar q $ systems. The results for $ cc\bar q\bar q$, $bb\bar q\bar q $ and $ bc\bar q\bar q $ are shown in Figs.~\ref{fig:ccqq}, \ref{fig:bbqq} and \ref{fig:bcqq}, respectively. We only focus on the energy spectra below the $ M(1S)M'(2S) $ dimeson thresholds. We identify resonant states by varying complex scaling angles $ \theta $. We obtain a series of $ QQ^{(\prime)}\bar q\bar q $ bound and resonant states, which are labeled as $ T_{QQ^{(\prime)},I(J)}(M) $ in the following discussions, where $ M $ represents the mass of the state. The complex energies, proportions of different color configurations and rms radii of these states are summarized in Table~\ref{tab:QQqq_structure}.
	\begin{table*}[htbp]
		\centering
		\caption{The complex energies $E=M-i \Gamma / 2$ (in MeV), proportions of different color configurations and rms radii (in fm) of the $Q_1Q_2 \bar{q}_1 \bar{q}_2$ bound and resonant states. The fourth column shows the binding energies $ \Delta E $ of the bound states. The last column shows the spatial configurations of the states, where C.E., C.DA., C.DC. and M. represent  compact even tetraquark, compact diquark-antidiquark tetraquark, compact diquark-centered tetraquark and molecular configurations, respectively. The constituent mesons of the molecular states are shown in the parentheses. The ``?" indicates that the rms radii results are numerically unstable as the complex scaling angle $ \theta $ varies. }
		\label{tab:QQqq_structure}
		\begin{tabular*}{\hsize}{@{}@{\extracolsep{\fill}}ccccccccccccc@{}}
			\hline\hline
			System&$ I(J^P)$& $ M-i\Gamma/2 $ & $ \Delta E $ & $ \chi_{\bar{3}_c\otimes3_c} $ &$ \chi_{6_c\otimes \bar6_c} $& $ r_{Q_1\bar{q}_1}^{\mathrm{rms}} $&$ r_{Q_2\bar{q}_2}^{\mathrm{rms}} $&$ r_{Q_1\bar{q}_2}^{\mathrm{rms}}$& $ r_{Q_2\bar{q}_1}^{\mathrm{rms}} $&$ r_{Q_1Q_2}^{\mathrm{rms}}$&$ r_{\bar{q}_1\bar{q}_2}^{\mathrm{rms}} $ & Configuration\\
			\hline
			$ cc\bar q\bar q $&$ 0(1^+) $&$3864$ &$ -14 $& $58\%$ & $42\%$ & $0.71$ & $0.64$ & $1.13$ & $1.16$ & $1.02$ & $1.22$&M.$ \,(D^*D) $\\
			&&	$ 4031-27i $&& $ 0\% $& $ 100\% $& $ 0.71 $ &$ 0.75 $ &?&?&?&?&?\\
			&&$4466-4i$ && $91\%$ & $9\%$ & $1.12$ & $1.09$ & $1.12$ & $1.13$ & $0.63$ & $0.86$&C.DA.\\
			&&$4542-10i$ && $62\%$ & $38\%$ & $0.95$ & $0.92$ & $1.02$ & $1.07$ & $0.89$ & $1.18$&C.E.\\
			&$ 1(2^{+}) $&$4673-2i$ && $90\%$ & $10\%$ & $1.15$ & $1.15$ & $1.19$ & $1.19$ & $0.68$ & $1.08$&C.DC.\\
			\hline
			$ bb\bar q\bar q $&$ 0(0^+) $&$11195$& $ -111 $ & $99\%$ & $1\%$ & $0.94$ & $0.94$ & $0.93$ & $0.93$ & $0.39$ & $1.08$&C.DC.\\
			&$ 0(1^+) $&$10491$ & $ -153 $ & $97\%$ & $3\%$ & $0.68$ & $0.67$ & $0.70$ & $0.71$ & $0.33$ & $0.78$&C.DC.\\
			&&$10642$ & $ -1 $ & $33\%$ & $67\%$ & $0.66$ & $0.63$ & $2.06$ & $2.07$ & $1.98$ & $2.15$&M. $ \,(\bar B^*\bar B) $\\
			&&$10700-1i$ && $44\%$ & $56\%$ & $0.67$ & $0.67$ & $1.96$ & $1.96$ & $1.88$ & $2.02$&M. $ \,(\bar B^*\bar B^*) $\\
			&&$11025-1i$ && $98\%$ & $2\%$ & $1.08$ & $1.07$ & $1.08$ & $1.08$ & $0.33$ & $0.83$&C.DA.\\
			&&$11152-0.5i$ && $60\%$ & $40\%$ & $0.89$ & $0.88$ & $0.96$ & $0.97$ & $0.77$ & $1.07$&C.E.\\
			&&$11249-0.3i$ && $40\%$ & $60\%$ & $0.91$ & $0.89$ & $1.12$ & $1.14$ & $0.75$ & $1.49$&C.E.\\
			&&$11291-1i$ && $77\%$ & $23\%$ & $1.06$ & $0.98$ & $1.10$ & $1.15$ & $0.76$ & $1.20$&C.E.\\
			&&$11315-1i$ && $82\%$ & $18\%$ & $1.00$ & $0.98$ & $1.05$ & $1.07$ & $0.85$ & $1.07$&C.E.\\
			&$ 0(2^+) $&$11370$ &$ -20 $& $91\%$ & $9\%$ & $1.04$ & $1.04$ & $1.09$ & $1.09$ & $0.41$ & $1.47$&C.DC.\\
			
			&$ 1(0^+) $&$11200-1i$ && $88\%$ & $12\%$ & $1.08$ & $1.08$ & $1.08$ & $1.08$ & $0.42$ & $1.03$&C.DC.\\
			&&$11262-4i$ && $57\%$ & $43\%$ & $0.91$ & $0.91$ & $0.99$ & $0.99$ & $0.63$ & $1.25$&C.E.\\
			
			&$ 1(1^+) $
			&$10685-7i$ && $91\%$ & $9\%$ & $0.73$ & $0.71$ & $0.74$ & $0.76$ & $0.31$ & $0.94$&C.DC.\\
			&&$11210-0.1i$ && $90\%$ & $10\%$ & $1.10$ & $1.09$ & $1.10$ & $1.10$ & $0.40$ & $1.02$&C.DC.\\
			
			&&$11286-5i$ && $63\%$ & $37\%$ & $0.96$ & $0.92$ & $0.95$ & $0.98$ & $0.62$ & $1.14$&C.E.\\
			
			&&$11312$ && $98\%$ & $2\%$ & $0.99$ & $1.01$ & $0.99$ & $1.00$ & $0.43$ & $1.16$&C.DC.\\
			
			&$ 1(2^+) $&$10715-2i$ && $90\%$ & $10\%$ & $0.74$ & $0.74$ & $0.74$ & $0.74$ & $0.37$ & $0.93$&C.DC.\\
			
			&&$11227-0.1i$ && $93\%$ & $7\%$ & $1.11$ & $1.11$ & $1.12$ & $1.12$ & $0.38$ & $1.02$&C.DC.\\
			
			&&$11305-5i$ && $59\%$ & $41\%$ & $0.94$ & $0.94$ & $0.97$ & $0.97$ & $0.66$ & $1.16$&C.E.\\
			\hline
			$ bc\bar q\bar q $&$ 0(0^+) $&$7129$ &$ -26 $& $48\%$ & $52\%$ & $0.64$ & $0.64$ & $0.91$ & $0.95$ & $0.76$ & $1.03$&C.E.\\
			&&$7301-73i$ && $31\%$ & $69\%$ & $0.70$ & $0.75$ & $0.79$ & $0.81$ & $0.43$ & $0.89$&C.DC.\\
			&&$7750-4i$ && $94\%$ & $6\%$ & $1.08$ & $1.08$ & $1.08$ & $1.11$ & $0.46$ & $0.85$&C.DA.\\
			&&$7843-9i$ && $63\%$ & $37\%$ & $0.75$ & $1.05$ & $0.90$ & $1.04$ & $0.82$ & $1.06$&C.E.\\
			&$ 0(1^+) $&$7185$ &$ -27 $& $60\%$ & $40\%$ & $0.67$ & $0.66$ & $0.88$ & $0.93$ & $0.71$ & $1.00$&C.E.\\
			&&$7764-2i$ && $95\%$ & $5\%$ & $1.09$ & $1.09$ & $1.09$ & $1.11$ & $0.47$ & $0.84$&C.DA.\\
			&&$7855-9i$ && $69\%$ & $31\%$ & $0.76$ & $1.06$ & $0.92$ & $1.04$ & $0.82$ & $1.05$&C.E.\\
			&$ 0(2^+) $&$7363$ &$ -3 $& $27\%$ & $73\%$ & $0.66$ & $0.70$ & $1.95$ & $1.97$ & $1.86$ & $2.05$&M. $ \,(\bar B^*D^*) $\\
			&&$7975$ && $26\%$ & $74\%$ & $0.75$ & $1.11$ & $1.18$ & $1.07$ & $0.86$ & $1.42$&C.E.\\
			&$ 1(2^+) $&$7430-28i$ && $93\%$ & $7\%$ & $0.74$ & $0.77$ & $0.75$ & $0.76$ & $0.40$ & $0.94$&C.DC.\\
			&&$7965-1i$ && $92\%$ & $8\%$ & $1.13$ & $1.12$ & $1.14$ & $1.14$ & $0.53$ & $1.05$&C.DC.\\
			&&$8025-2i$ && $80\%$ & $20\%$ & $0.80$ & $1.12$ & $0.95$ & $1.03$ & $0.73$ & $1.10$&C.E.\\
			\hline\hline
		\end{tabular*}
	\end{table*}
	\subsubsection{$ cc\bar q\bar q $}
	
	For the $ cc\bar q\bar q $ system, we obtain one bound state, three resonant states with quantum number $ I(J^P)=0(1^+) $, and a resonant state with $ I(J^P)=1(2^+) $. The isoscalar bound state $ T_{cc,0(1)}(3864) $ is located $ 14 $ MeV below the $ D^*D $ threshold. It has a molecular configuration, with $ r_{c_1\bar q_1}^{\rm rms} $ and $ r_{c_2\bar q_2}^{\rm rms} $ close to the sizes of $ D^* $ and $ D $ mesons, respectively, indicating that it is a $ D^*D $ molecular state. Comparing with the experimentally observed $ T_{cc}(3875)^+ $, whose binding energy is around $ -300 $ keV and characteristic size is around $ 7 $ fm~\cite{LHCb:2021auc}, $ T_{cc,0(1)}(3864) $ is lower in energy and smaller in size. However, considering that the uncertainties of quark model are up to tens of MeV, it may still serve as a candidate for $ T_{cc}(3875)^+ $. The discrepancies suggest that improvements of the constituent quark model are needed to better describe the long-range interactions between hadrons. Additionally, in Sec.~\ref{sec:varymQ}, we calculate the rms radii of $T_{QQ,0(1)}$ with different binding energies by varying the heavy quark mass. When the binding energy is adjusted to match that of the experimental $T_{cc}(3875)^+$, the distance between the (anti)quarks in $D$ and $D^*$ is approximately $ 6 $ fm. This result is in excellent agreement with the characteristic size estimated in Ref.~\cite{LHCb:2021auc}, further supporting the interpretation of the $T_{cc}(3875)^+$ as a very loosely bound molecular state.
	
	We obtain an isoscalar resonant state $ T_{cc,0(1)}(4031) $ with a width of $ \Gamma=54 $ MeV, located $ 1 $ MeV below the $ D^*D^* $ threshold. This state is in agreement with the results from previous works using heavy quark spin symmetry~\cite{Albaladejo:2021vln}, constituent quark model~\cite{Meng:2024yhu} and lattice QCD~\cite{Whyte:2024ihh}. It can be searched for in the $ D^*D $ and $ DD\pi $ channel in future experiments. One may expect that $ T_{cc,0(1)}(4031) $ is the partner of the $ D^*D $ molecular bound state $ T_{cc,0(1)}(3864) $ and has a $ D^*D^* $ molecular configuration. However, the rms radii for $ T_{cc,0(1)}(4031) $ is highly numerically unstable and changes dramatically as $ \theta $ varies. The numerical instability of the rms radii may result from the state being sandwiched between the $ D^*D $ and $ D^*D^* $ thresholds, which makes it strongly coupled to the scattering states. Calculations with higher numerical precision are needed to identify its spatial structure.
	
	We also obtain three higher resonant states $ T_{cc,0(1)}(4466), T_{cc,0(1)}(4542), $ and $ T_{cc,1(2)}(4673) $, which have compact tetraquark configurations.  The state $ T_{cc,1(2)}(4673) $ has 10\% isovector $(\bar q\bar q)_{6_c}^1$ component, where the spatial wave function between two light antiquarks must be antisymmetric. Similar configurations also appear in other $2^+$ $QQ^{(\prime)}\bar q\bar q$ states. Antisymmetric spatial wave function is possible in our S-wave calculations since we only restrict the total orbital angular momentum to be S wave. The states $ T_{cc,0(1)}(4466)$ and $ T_{cc,0(1)}(4542) $ can decay into the $ D^*D $ and $ D^*D^* $ channels while the isovector state can only decay into the $ D^*D^* $ channel. These three states were not found in Ref.~\cite{Meng:2024yhu}, where the authors used the same constituent quark model as we do. The reason for the discrepancies may be that we use a larger set of wave function bases compared to the previous work, as discussed in Ref.~\cite{Wu:2024euj}. Future experimental explorations of these states may help resolve the discrepancies between different methods.
	
	\subsubsection{$ bb\bar q\bar q $}
	We first focus on the $ I(J^P)=0(1^+) $ $ bb\bar q\bar q $ system, which is the bottom analog of the $ T_{cc} $ state. We obtain a deeply bound state $ T_{bb,0(1)}(10491) $ with a binding energy of $ -153 $ MeV, and a shallow bound state $ T_{bb,0(1)}(10642) $ with a binding energy of $ -1 $ MeV. The lower state is a compact diquark-centered tetraquark, which is similar to the helium atom. The two bottom quarks form a very compact diquark cluster, with the rms radius between them ($ r_{b_1b_2}^{\rm rms}=0.33 $ fm) being approximately the size of the bottomonia, as listed in Table~\ref{tab:bbrms}. The two light antiquarks orbit around the diquark, similar to the way two electrons orbit around the nucleus in a helium atom. The rms radii between the bottom quarks and light antiquarks $ r_{b\bar q}^{\rm rms} $ are around $ 0.7 $ fm. The dominant color configuration of the state is $ \chi_{\bar 3_c\otimes 3_c} $, where the color electric interactions are attractive between two (anti)quarks. Note that the $ \chi_{\bar 3_c\otimes 3_c} $ component reaches $97\%$! The strong attraction between two bottom quarks contributes to the deep binding energy of the state.  A deeply bound state in the $ I(J^P)=0(1^+) $ $ bb\bar q\bar q $ system has been anticipated since the 1980s~\cite{Zouzou:1986qh} and was recently predicted in lattice QCD studies~\cite{Junnarkar:2018twb,Aoki:2023nzp,Colquhoun:2024jzh}. Our findings agree with the previous works. The bound state $ T_{bb,0(1)}(10491) $ is below the $ \bar B\bar B $ threshold and can only decay weakly. 
	
	The higher bound state $ T_{bb,0(1)}(10642) $ has a molecular configuration, with $ r_{b_1\bar q_1}^{\rm rms} $ and $ r_{b_2\bar q_2}^{\rm rms} $ being close to the sizes of $ \bar B^* $ and $ \bar B $ mesons, respectively, indicating that it is a $ \bar B^*\bar B $ molecular state. It is the bottom analog of the $ D^*D $ molecular bound state $ T_{cc,0(1)}(3864) $. The bound state $ T_{bb,0(1)}(10642) $ can decay radiatively to $ \bar B\bar B\gamma $.
	
	\begin{table}
		\centering
		\caption{The rms radii (in fm) of bottomonia.}
		\label{tab:bbrms}
		\begin{tabular}{cc|cc}
			\hline\hline
			Mesons&$ r^{\rm rms}_{\rm Theo.} $&	Mesons&$ r^{\rm rms}_{\rm Theo.} $ \\
			\hline
			$ \eta_b $&$ 0.20 $&$ \eta_b(2S) $&$ 0.48 $\\
			$ \Upsilon $&$ 0.21 $&$ \Upsilon(2S) $&$ 0.49 $\\
			\hline\hline
		\end{tabular}
	\end{table}

	In addition to bound states, we also obtain six isoscalar $ bb\bar q\bar q $ resonant states with $ J^P=1^+ $. The state $ T_{bb,0(1)}(10700) $ is located near the $ \bar B^*\bar B^* $ threshold and identified as a $ \bar B^*\bar B^* $ molecular state. It is dominated by the $ \bar B^*\bar B^* $ component, resulting in a small decay width of only $2$ MeV to the $ \bar B^*\bar B $ channel. In contrast, its charmed partner $ T_{cc,0(1)}(4031) $ is a relatively broad state with a width of $54$ MeV. This discrepancy may indicate that the nature of the two states is different, namely $ T_{cc,0(1)}(4031) $ may have sizable contributions from both the $ D^*D $ and $ D^*D^* $ channels. The state $ T_{bb,0(1)}(10700) $ can be searched for in the $ \bar B^*\bar B $ channel.
	
	The state $ T_{bb,0(1)}(11025) $ is a compact diquark-antidiquark tetraquark state. Comparing its internal structures with those of the bound state $ T_{bb,0(1)}(10491) $, we find an interesting resemblance. Both states are dominated by the $ \chi_{\bar 3_c\otimes3_c} $ color configuration. The rms radii between two bottom quarks $ r^{\rm rms}_{b_1b_2}=0.33 $ fm are the same for these two states, while the rms radii between the bottom quarks and light antiquarks $ r^{\rm rms}_{b\bar q} $ are around $ 1 $ fm for $ T_{bb,0(1)}(11025) $, larger than those for $ T_{bb,0(1)}(10491) $. This clearly suggests that $ T_{bb,0(1)}(11025) $ is the radial excitation in the light degree of freedom of $ T_{bb,0(1)}(10491) $. The state $ T_{bb,0(1)}(11025) $ can decay into the $ \bar B^*\bar B $ and $ \bar B^*\bar B^* $ channels.
	
	Two isoscalar bound states $ T_{bb,0(0)}(11195) $ and $ T_{bb,0(2)}(11370) $ are found in the $ 0^+ $ and $ 2^+ $ systems, respectively. The $ T_{bb,0(0)}(11195) $ does not decay into the S-wave $BB$ pair and $ T_{bb,0(2)}(11370) $ does not decay into the S-wave $B^*B^*$ pair, since the total wave functions of two identical bosons must be interchange symmetric. Hence, their lowest S-wave dimeson thresholds are $M(1S) M'(2S)$. However, it should be noted that P-wave dimeson channels exist below these states. The coupling between the P-wave channels and these states may alter their positions and nature.
	
	In contrast to the isoscalar systems, no bound state is obtained in the isovector $ bb\bar q\bar q $ systems. The absence of the isovector bound state suggests that the configuration $ (\bar q\bar q)^0_{3_c} $ with isospin 0, which is referred to as ``good antidiquark" in Ref.~\cite{Jaffe:2004ph,Deng:2022cld}, is essential for forming the doubly bottomed tetraquark bound states. We obtain two resonant states with $ J^P=0^+ $, four resonant states with $ J^P=1^+ $, and three resonant states with $ J^P=2^+ $. They are identified as either compact diquark-centered tetraquarks or compact even tetraquarks. The two lowest states $ T_{bb,1(1)}(10685) $ and $ T_{bb,1(2)}(10715) $ may be more likely to be observed in experiments. It is worth noting that in Ref.~\cite{Meng:2023for}, the authors used a different constituent quark model including the one-pion exchange interaction, and identified an isovector resonant state with $ J^P=1^+ $, whose mass and width are in agreement with the state $ T_{bb,1(1)}(10685) $ obtained in our calculations. We recommend experimental exploration of this state in the $ \bar B^*\bar B $ channel.

	\subsubsection{$ bc\bar q\bar q $}
	We identify three isoscalar bound states $ T_{bc,0(0)}(7129)$, $T_{bc,0(1)}(7185) $, and $ T_{bc,0(2)}(7363) $ in the $ bc\bar q\bar q $ systems. $ T_{bc,0(0)}(7129)$ and $ T_{bc,0(1)}(7185) $ are compact even tetraquark states, where the rms radii between four (anti)quarks are of similar size. Unlike the $ \chi_{\bar{3}_c\otimes3_c} $ dominated compact diquark-centered tetraquark $ T_{bb,0(1)}(10491) $, these two compact even tetraquarks have sizable components of both $ \chi_{\bar{3}_c\otimes3_c} $ and $ \chi_{6_c\otimes \bar6_c} $ configurations. This suggests that both the interactions between two (anti)quarks and the interactions between quarks and antiquarks are important for these two states. The shallow bound state $ T_{bc,0(2)}(7363) $ with a binding energy of $ -3 $ MeV is identified as a $ \bar B^*D^* $ molecular state. The scalar bound state $ T_{bc,0(0)}(7129)$ can only decay weakly, while $ T_{bc,0(1)}(7185) $ can decay radiatively to $ \bar B D\gamma $, and $ T_{bc,0(2)}(7363) $ can decay strongly to $ \bar B^*D\pi $. In contrast to the isoscalar case, no isovector $ bc\bar q\bar q $ bound state is found. This again shows that the $ (\bar q\bar q)^0_{3_c} $ configuration plays a vital role in forming doubly heavy tetraquark bound states, as in the $ bb\bar q\bar q $ systems. In Refs.~\cite{Li:2012ss,Sakai:2017avl}, the authors investigated possible $ bc\bar q\bar q $ bound states from the molecular picture. It was shown in  Ref.~\cite{Sakai:2017avl} that for the isovector system the interactions between charmed and bottomed mesons are repulsive and no bound state exists, which is consistent with our findings. Moreover, both studies~\cite{Li:2012ss,Sakai:2017avl} suggested that shallow bound states with binding energies of several MeV may exist in the $ I(J^P)=0(0^+),0(1^+), $ and $ 0(2^+) $ $ bc\bar q\bar q $ systems. For comparison, we also obtain a loosely bound molecular state in the isoscalar $ 2^+ $ system. But for the isoscalar $ 0^+ $ and $ 1^+ $ systems, our quark model calculations lead to $ bc\bar q\bar q $  compact tetraquark states with binding energies around $ -30 $ MeV. 
	
	We also obtain a series of resonant states with different quantum numbers. The two lowest states $ T_{bc,0(0)}(7301) $ and $ T_{bc,1(2)}(7430) $ may be more likely to be observed in experiments. They can be searched for in the $ \bar B D $ and $ \bar B^*D^* $ channels, respectively. Moreover, in the $ I(J^P)=0(1^+)$ $ bc\bar q\bar q$ system, we find that the scattering states of $ \bar BD^* $ and $ \bar B^*D^* $ deviate from the continuum lines, indicating the possible existence of a resonant state, which may be the $ bc\bar q\bar q $ analog of $ T_{cc,0(1)}(4031) $ and $ T_{bb,0(1)}(10700) $. However, we cannot determine this state accurately due to the limitations of numerical precision in the present calculations. 
	
	\subsection{$ QQ^{(\prime)}\bar s\bar q $}
	\begin{table*}[htbp]
		\centering
		\caption{The complex energies $E=M-i \Gamma / 2$ (in MeV), proportions of different color configurations and the rms radii (in fm) of the $Q_1Q_2 \bar{s} \bar{q}$ bound and resonant states. The fourth column shows the binding energies $ \Delta E $ of the bound states. The last column shows the spatial configurations of the states, where C.E., C.DA. and C.DC. represent  compact even tetraquark, compact diquark-antidiquark tetraquark and compact diquark-centered tetraquark, respectively. The ``?" indicates that the rms radii results are numerically unstable as the complex scaling angle $ \theta $ varies.   }
		\label{tab:QQsq_structure}
		\begin{tabular*}{\hsize}{@{}@{\extracolsep{\fill}}ccccccccccccc@{}}
			\hline\hline
			System&$ J^P$& $ M-i\Gamma/2 $ & $ \Delta E $ & $ \chi_{\bar{3}_c\otimes3_c} $ &$ \chi_{6_c\otimes \bar6_c} $& $ r_{Q_1\bar{s}}^{\mathrm{rms}} $&$ r_{Q_2\bar{q}}^{\mathrm{rms}} $&$ r_{Q_1\bar{q}}^{\mathrm{rms}}$& $ r_{Q_2\bar{s}}^{\mathrm{rms}} $&$ r_{Q_1Q_2}^{\mathrm{rms}}$&$ r_{\bar{s}\bar{q}}^{\mathrm{rms}} $&Configuration\\
			\hline
			$ cc\bar s\bar q $&$ 1^+ $&$4578-3i$ && $74\%$ & $26\%$ & $1.02$ & $0.90$ & $1.26$ & $1.29$ & $1.03$ & $1.23$	&C.E. \\
			&$ 2^+ $&$4723-2i$ && $81\%$ & $19\%$ & $1.07$ & $0.95$ & $1.17$ & $1.24$ & $0.83$ & $1.19$&C.E.\\
			\hline
			$ bb\bar s\bar q $&$ 0^+ $&$11243-1i$ && $89\%$ & $11\%$ & $0.98$ & $0.91$ & $0.95$ & $0.99$ & $0.41$ & $0.98$&C.DC. \\
			&&$11280-1i$ && $90\%$ & $10\%$ & $0.84$ & $0.84$ & $0.88$ & $0.84$ & $0.45$ & $1.00$&C.DC.\\
			&$ 1^+ $&$10647$ &$ -64 $& $91\%$ & $9\%$ & $0.56$ & $0.67$ & $0.71$ & $0.61$ & $0.36$ & $0.76$&C.DC.\\
			&&$10766-16i$ && $92\%$ & $8\%$ & $0.58$ & $0.71$ & $0.74$ & $0.63$ & $0.36$ & $0.85$&C.DC.\\
			&&$11138-0.5i$ && $97\%$ & $3\%$ & $0.95$ & $0.97$ & $0.98$ & $0.96$ & $0.34$ & $0.80$&C.DA.\\
			&&$11254-0.1i$ && $93\%$ & $7\%$ & $0.99$ & $0.94$ & $0.97$ & $1.00$ & $0.39$ & $0.96$&C.DC.\\
			&&$11256-5i$ && $60\%$ & $40\%$ &  $0.77$ & $0.84$ & $0.91$ & $0.82$ & $0.71$ & $0.99$&C.E.\\
			
			&$ 2^+ $&$10798-8i$ && $90\%$ & $10\%$ & $0.59$ & $0.72$ & $0.77$ & $0.66$ & $0.39$ & $0.88$&C.DC.\\
			&&$11272$ && $95\%$ & $5\%$ & $1.00$ & $0.96$ & $0.98$ & $1.00$ & $0.37$ & $0.95$&C.DC.\\
			&&$11360-5i$ && $62\%$ & $38\%$ & $0.80$ & $0.87$ & $0.88$ & $0.83$ & $0.62$ & $1.04$&C.E.\\
			\hline
			$ bc\bar s\bar q $&$ 0^+ $&$7420-19i$ && $61\%$ & $39\%$ & $0.63$ & $0.78$ & $0.77$ & $0.71$ & $0.60$ & $0.90$&C.E. \\
			&&$7853-3i$ && $74\%$ & $26\%$ & $0.97$ & $0.87$ & $1.12$  & $1.19$ & $0.84$ & $1.17$&C.E.\\
			
			&$ 1^+ $&$7437-10i$ && $24\%$ & $76\%$ & ? & ? & ? & ? & ? & ?&?\\
			
			&&$7872-2i$ && $83\%$ & $17\%$ & $0.97$ & $0.92$ & $1.04$ & $1.09$ & $0.68$ & $1.02$&C.E. \\
			&$ 2^+ $&$7518-45i$ && $100\%$ & $0\%$ & $0.61$ & $0.75$ & $0.72$ & $0.63$ & $0.37$ & $0.84$&C.DC.\\
			&&$8010-1i$ && $88\%$ & $12\%$ & $1.04$ & $0.97$ & $0.99$ & $1.05$ & $0.58$ & $1.06$&C.DC.\\	
			\hline\hline
		\end{tabular*}
	\end{table*}

	\begin{table*}[htbp]
		\centering
		\caption{ The complex energies $E=M-i \Gamma / 2$ (in MeV), proportions of different color configurations and the rms radii (in fm) of the $Q_1Q_2 \bar{s} \bar{s}$ resonant states. The last column shows the spatial configurations of the states, where C.E. and C.DC. represent compact even tetraquark and compact diquark-centered tetraquark, respectively. }
		\label{tab:QQss_structure}
		\begin{tabular*}{\hsize}{@{}@{\extracolsep{\fill}}cccccccccccc@{}}
			\hline\hline
			System&$ J^P$& $ M-i\Gamma/2 $  & $ \chi_{\bar{3}_c\otimes3_c} $ &$ \chi_{6_c\otimes \bar6_c} $& $ r_{Q_1\bar{s}_1}^{\mathrm{rms}} $&$ r_{Q_2\bar{s}_2}^{\mathrm{rms}} $&$ r_{Q_1\bar{s}_2}^{\mathrm{rms}}$& $ r_{Q_2\bar{s}_1}^{\mathrm{rms}} $&$ r_{Q_1Q_2}^{\mathrm{rms}}$&$ r_{\bar{s}_1\bar{s}_2}^{\mathrm{rms}} $&Configuration\\
			\hline
			$ cc\bar s\bar s $&$2^{+}$ & $4808 - 4 i$ & $83\%$ & $17\%$ & $0.94$ & $0.94$ & $1.11$ & $1.11$ & $0.80$ & $1.06$& C.E. \\
			\hline
			$ bb\bar s\bar s $&	$0^{+}$ & $11317 - 2 i$ & $89\%$ & $11\%$ & $0.87$ & $0.87$& $0.89$ & $0.89$ & $0.43$ & $0.86$ &C.DC.\\
			&$1^{+}$& $10846 - 29 i$ & $96\%$ & $4\%$  & $0.58$ & $0.56$ & $0.58$ & $0.60$& $0.30$ & $0.73$&C.DC. \\
			&& $11328 - 1i$ & $92\%$ & $8\%$  & $0.90$ & $0.86$ & $0.89$ & $0.91$ & $0.40$ & $0.84$&C.DC. \\
			&$2^{+}$ & $10877 - 16 i$ & $93\%$ & $7\%$ & $0.58$ & $0.58$& $0.62$ & $0.62$ & $0.34$ & $0.75$ &C.DC.\\ 
			&&$11346$ & $94\%$ & $6\%$  & $0.89$ & $0.89$ & $0.91$ & $0.91$& $0.38$ & $0.83$&C.DC. \\
			&& $11430 - 4 i$ & $66\%$ & $34\%$ & $0.75$ & $0.75$ & $0.78$ & $0.78$& $0.59$ & $0.94$&C.E. \\
			\hline
			$ bc\bar s\bar s $&$2^{+}$ & $7605 - 64 i$ & $108\%$ & $-8\%$ & $0.58$ & $0.63$ & $0.62$& $0.66$ & $0.40$ & $0.75$& C.DC. \\
			&& $8090 - 2 i$ & $91\%$ & $9\%$  & $0.92$ & $0.91$ & $0.96$ & $0.97$ & $0.55$ & $0.90$&C.DC. \\
			\hline\hline
		\end{tabular*}
	\end{table*}

\begin{figure*}[htbp]
		\centering
		\includegraphics[width=1\linewidth]{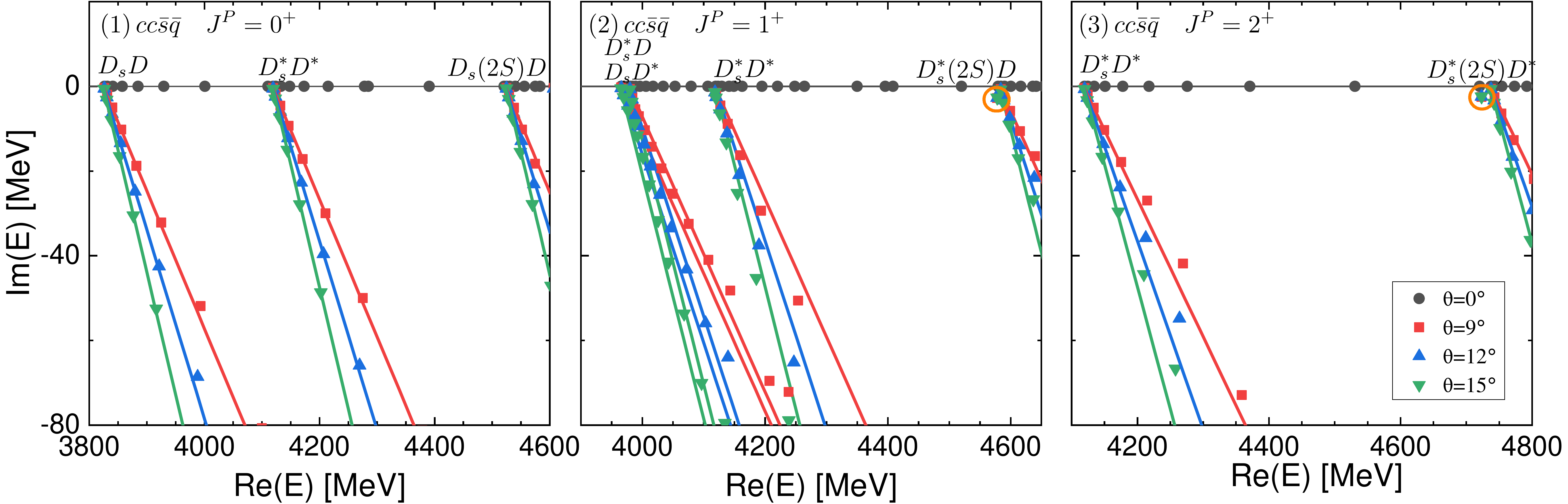}
		\caption{The complex energy eigenvalues of the  $cc\bar s\bar q$ states with varying $\theta$ in the CSM. The solid lines represent the continuum lines rotating along $\operatorname{Arg}(E)=-2 \theta$. The resonant states do not shift as $\theta$ changes and are highlighted by the orange circles.}
		\label{fig:ccsq}
	\end{figure*}

\begin{figure*}[htbp]
		\centering
		\includegraphics[width=1\linewidth]{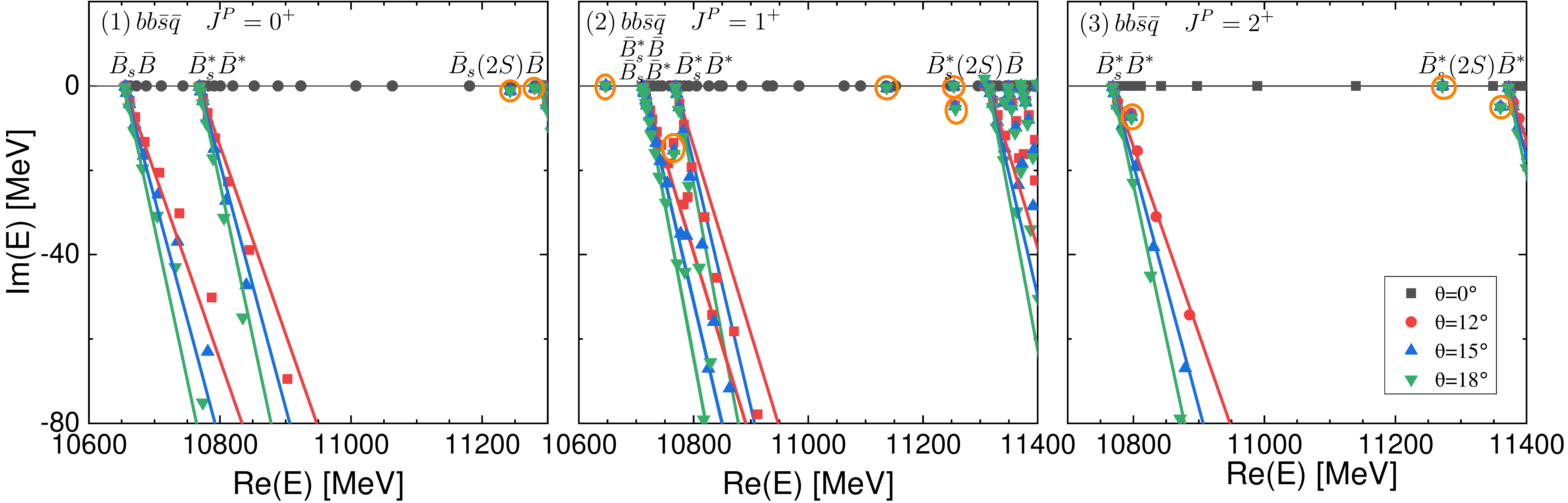}
		\caption{The complex energy eigenvalues of the  $bb\bar s\bar q$ states with varying $\theta$ in the CSM. The solid lines represent the continuum lines rotating along $\operatorname{Arg}(E)=-2 \theta$. The bound and resonant states do not shift as $\theta$ changes and are highlighted by the orange circles.}
		\label{fig:bbsq}
	\end{figure*}

\begin{figure*}[htbp]
		\centering
		\includegraphics[width=1\linewidth]{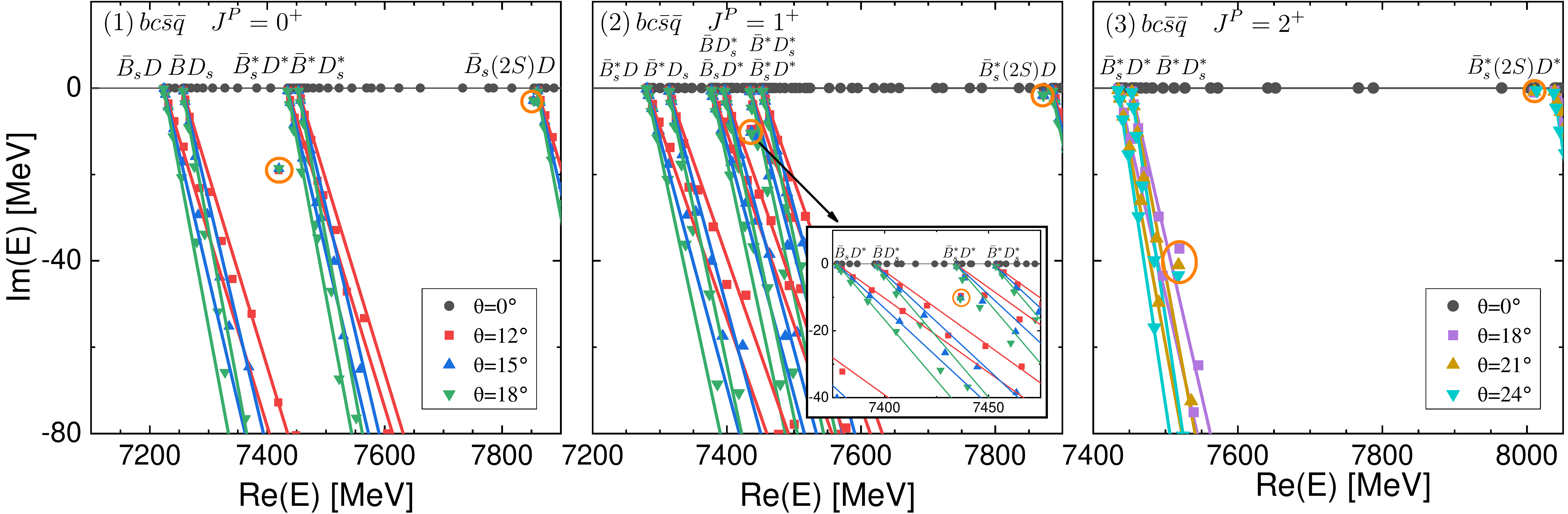}
		\caption{The complex energy eigenvalues of the  $bc\bar s\bar q$ states with varying $\theta$ in the CSM. The solid lines represent the continuum lines rotating along $\operatorname{Arg}(E)=-2 \theta$. The resonant states do not shift as $\theta$ changes and are highlighted by the orange circles.}
		\label{fig:bcsq}
	\end{figure*}

	The complex eigenenergies of the S-wave $ cc\bar s\bar q$, $bb\bar s\bar q $, and $ bc\bar s\bar q $ systems are shown in Figs.~\ref{fig:ccsq}, \ref{fig:bbsq} and \ref{fig:bcsq}, respectively. We obtain a series of $ QQ^{(\prime)}\bar s\bar q $ bound and resonant states, which are labeled as $ T_{QQ^{(\prime)}\bar s,J}(M) $ in the following discussions, where $ M $ represents the mass of the state. The complex energies, proportions of different color configurations and rms radii of these states are summarized in Table~\ref{tab:QQsq_structure}.
	
	We obtain two resonant states in the $ cc\bar s\bar q $ system. They are located near the $ M(1S)M'(2S) $ dimeson threshold and have compact even tetraquark configurations. $ T_{cc\bar s,1}(4578) $ can decay into the $ D_s^*D$, $D_sD^*$, and $ D_s^*D^* $ channel while $ T_{cc\bar s,2}(4723) $ can only decay into the $ D_s^*D^* $ channel.
	
	For the $ bb\bar s\bar q $ system, we identify a bound state $ T_{bb\bar s,1}(10647) $ with a binding energy of $ -64 $ MeV. It is a deeply bound state with compact diquark-centered tetraquark configuration, and can be considered as the strange partner of $ T_{bb,0(1)}(10491) $. It is below the $ \bar B_s\bar B $ threshold and can only decay weakly. Moreover, we also obtain four resonant states with  $ J^P=1^+ $. The state $ T_{bb\bar s,1}(10766) $ is located $ 3 $ MeV below the $ \bar B_s^*\bar B^* $ threshold and have a compact diquark-centered tetraquark configuration. It is the radial excitation in the light degree of freedom of $T_{bb\bar s,1}(10647)$, similar to $T_{bb,0(1)}(11025)$ being the radial excitation of $T_{bb,0(1)}(10491)$. It can decay into the $ \bar B_s^*\bar B $ and $ \bar B_s\bar B^* $ channel, while the three higher resonant states can also decay into the $ \bar B_s^*\bar B^* $ channel. In addition, we obtain two resonant states with $ J^P=0^+ $ and three resonant states with $ J^P=2^+ $. The $ 0^+ $ states can decay into the $ \bar B_s\bar B $ and $ \bar B_s^*\bar B^* $ channels while the $ 2^+ $ states can only decay into the $ \bar B_s^*\bar B^* $ channel.
	
	For the $ bc\bar s\bar q $ system, we obtain two resonant states each in the $ 0^+, 1^+, $ and $ 2^+ $ systems. Similar to the resonant state $ T_{cc,0(1)}(4031) $, $ T_{bc\bar s,1}(7437) $ is sandwiched between two thresholds, and its rms radii results are numerically unstable. The lowest $ 0^+ $ resonant state can decay into the $ \bar B_sD $ and $ \bar BD_s $ channels. The lowest $ 1^+ $ resonant state can decay into the $ \bar B_s^*D, \bar B^*D_s, \bar B_sD^*, $ and $ \bar BD_s^* $ channels. The lowest $ 2^+ $ resonant state can decay into the $ 
	\bar B_s^*D^* $ and $ \bar B^*D_s^* $ channels.

	\subsection{$ QQ^{(\prime)}\bar s\bar s $}

\begin{figure*}[htbp]
		\centering
		\includegraphics[width=1\linewidth]{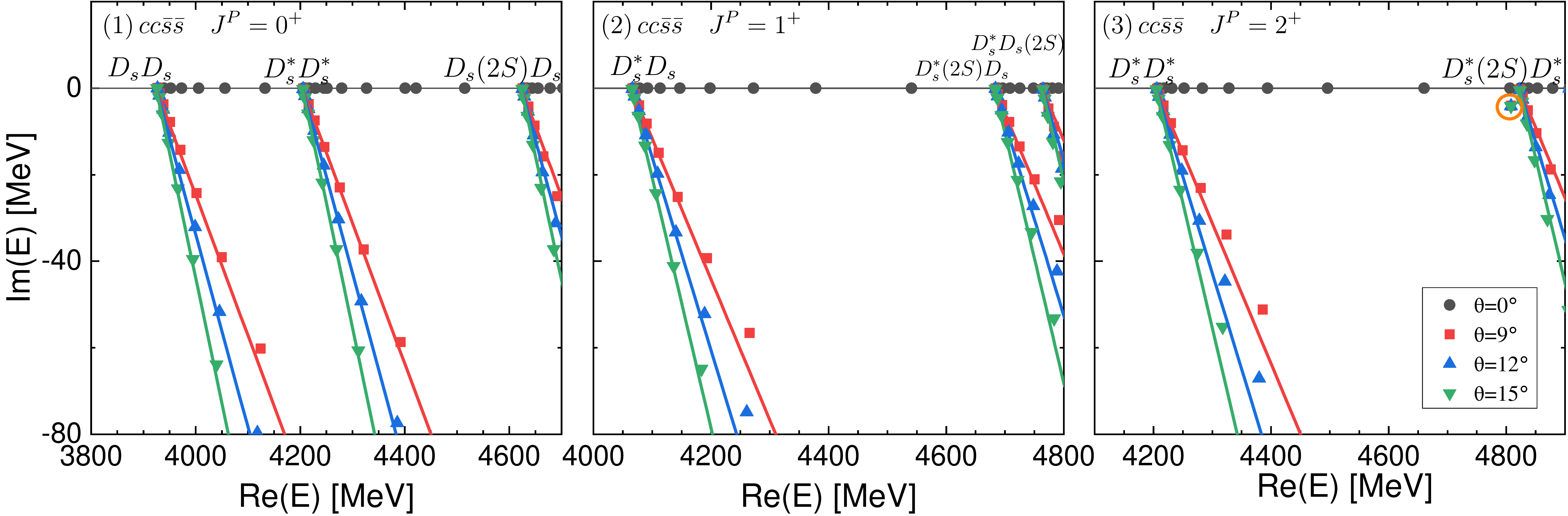}
		\caption{The complex energy eigenvalues of the  $cc\bar s\bar s$ states with varying $\theta$ in the CSM. The solid lines represent the continuum lines rotating along $\operatorname{Arg}(E)=-2 \theta$. The resonant states do not shift as $\theta$ changes and are highlighted by the orange circles.}
		\label{fig:ccss}
	\end{figure*}

\begin{figure*}[htbp]
		\centering
		\includegraphics[width=1\linewidth]{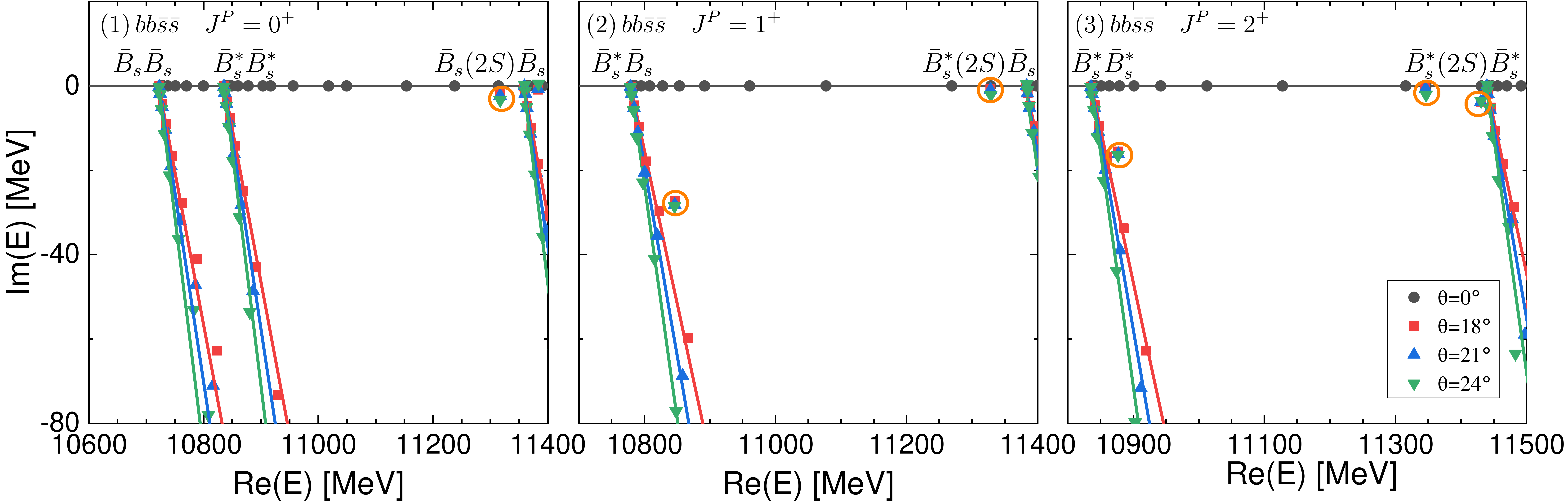}
		\caption{The complex energy eigenvalues of the  $bb\bar s\bar s$ states with varying $\theta$ in the CSM. The solid lines represent the continuum lines rotating along $\operatorname{Arg}(E)=-2 \theta$. The resonant states do not shift as $\theta$ changes and are highlighted by the orange circles.}
		\label{fig:bbss}
	\end{figure*}

\begin{figure*}[htbp]
		\centering
		\includegraphics[width=1\linewidth]{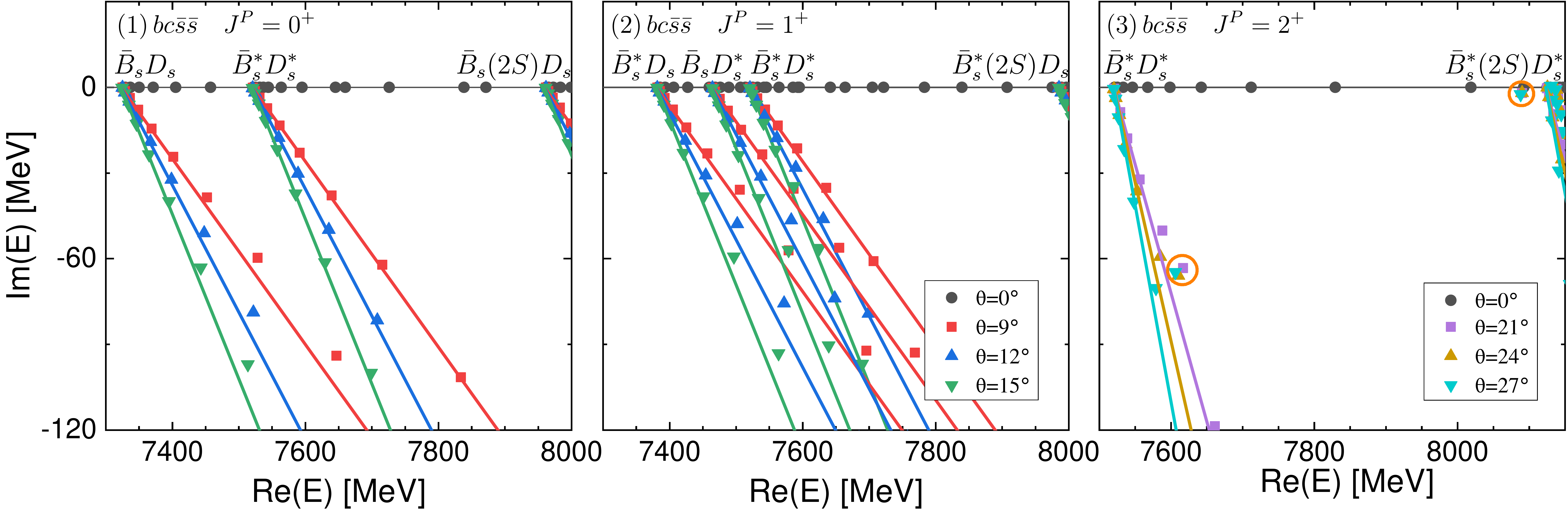}
		\caption{The complex energy eigenvalues of the  $bc\bar s\bar s$ states with varying $\theta$ in the CSM. The solid lines represent the continuum lines rotating along $\operatorname{Arg}(E)=-2 \theta$. The resonant states do not shift as $\theta$ changes and are highlighted by the orange circles.}
		\label{fig:bcss}
	\end{figure*}

	The complex eigenenergies of the S-wave $ cc\bar s\bar s, bb\bar s\bar s, $ and $ bc\bar s\bar s $ systems are shown in Figs.~\ref{fig:ccss}, \ref{fig:bbss} and \ref{fig:bcss}, respectively. We obtain a series of $ QQ^{(\prime)}\bar s\bar s $ resonant states, which are labeled as $ T_{QQ^{(\prime)}\bar s\bar s,J}(M) $ in the following discussions, where $ M $ represents the mass of the state. The complex energies, proportions of different color configurations and rms radii of these states are summarized in Table~\ref{tab:QQss_structure}.
	
	The $ QQ^{(\prime)}\bar s\bar s $ systems and the isovector $ QQ^{(\prime)}\bar q\bar q $ systems share the same internal symmetries, and their energy spectra bear a resemblance. In the $ cc\bar s\bar s $ system, we obtain a resonant state $ T_{cc\bar s\bar s,2}(4808) $ below the $ D_s^*(2S)D_s^* $ threshold. It can be considered as the strange partner of $ T_{cc,1(2)}(4673) $. In the $ bb\bar s\bar s $ system, the two lowest resonant states $ T_{bb\bar s\bar s,1}(10846) $ and $ T_{bb\bar s\bar s,2}(10877) $ are the strange partners of $ T_{bb,1(1)}(10685) $ and $ T_{bb,1(2)}(10715) $, respectively. In the $ bc\bar s\bar s $ system, two resonant states with $ J^P=2^+ $ are found. The lowest state $ T_{bc\bar s\bar s,2}(7605) $ is the strange partner of $ T_{bc,1(2)}(7430) $. The mass differences between the isovector $ QQ^{(\prime)}\bar q\bar q $ resonant states and their strange partners $ QQ^{(\prime)}\bar s\bar s $ resonant states are around $ 150 $ MeV. For higher resonant states, the correspondence between $ QQ^{(\prime)}\bar s\bar s $ states and isovector $ QQ^{(\prime)}\bar q\bar q $ states is less clear. Compared to the isovector $ QQ^{(\prime)}\bar q\bar q $ systems, the $ QQ^{(\prime)}\bar s\bar s $ systems have fewer resonant states near the $ M(1S)M'(2S) $ dimeson thresholds. All of the $ QQ^{(\prime)}\bar s\bar s $ states are identified as compact tetraquarks and can be searched for in corresponding dimeson decay channels.
	
	The state $ T_{bc\bar s\bar s,2}(7605) $ has a negative proportion of color configuration $ \chi_{6_c\otimes\bar 6_c} $, which is rather confusing. However, it should be emphasized that the inner products calculated by the c product in Eq.~\eqref{eq:cpro} are not positive definite and generally not real. Moreover, the physical quantities calculated by the c product may only have probabilistic interpretation for resonant states that are not too broad~\cite{Berggren:1970wto,homma1997matrix}. Therefore, the negative proportions of the broad resonance $ T_{bc\bar s\bar s,2}(7605) $ should not be taken too seriously.
	

	\subsection{Heavy quark mass dependence of $ T_{QQ,0(1)} $ bound states}~\label{sec:varymQ}
\begin{figure}[htbp]
		\centering
		\includegraphics[width=0.85\linewidth]{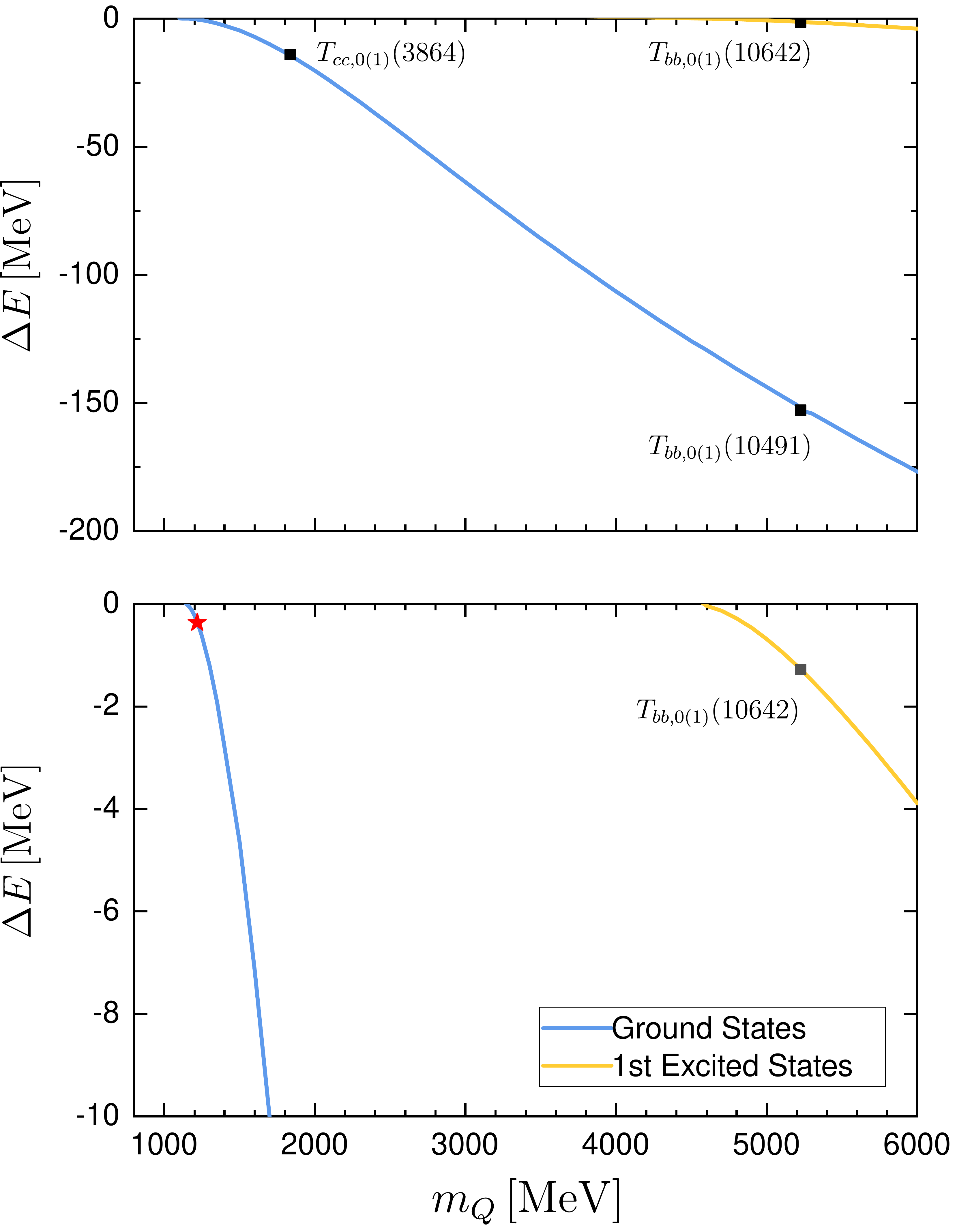}
		\caption{The binding energies $ \Delta E $ of the ground states and the first excited states in the $ I(J^P)=0(1^+) $ $ QQ\bar q\bar q $ systems with varying heavy quark mass $ m_Q $. The lower panel shows the energy region near the threshold more clearly. The physical points are indicated by black points. The red star point corresponds to the binding energy adjusted to match that of the experimental $ T_{cc}(3875)^+ $.}  
		\label{fig:evsm}
	\end{figure}

\begin{figure*}
		\centering
		\includegraphics[width=0.9\linewidth]{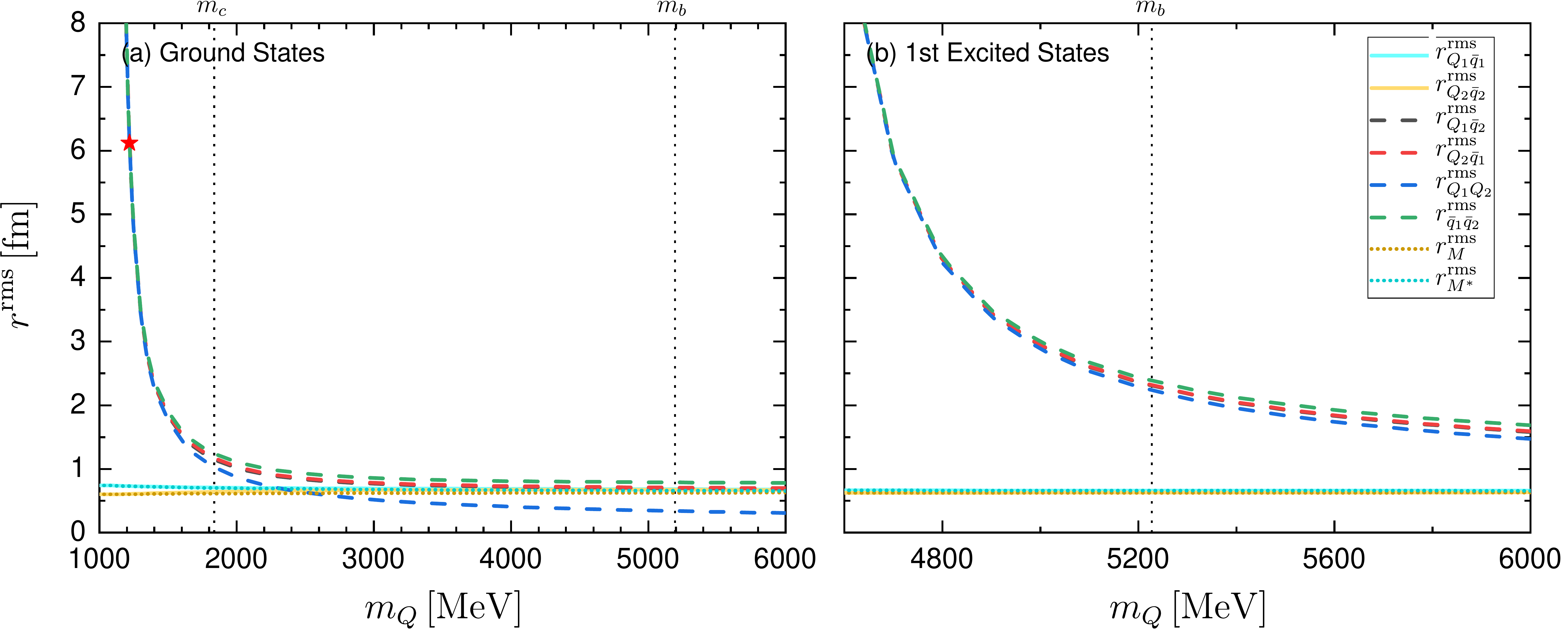}
		\caption{The rms radii (in fm) of (a) the ground states and (b) the first excited states in the $ I(J^P)=0(1^+) $ $ QQ\bar q\bar q $ systems with varying heavy quark mass $ m_Q $. The solid lines show the rms radii within the clusters $ Q_1\bar q_1 $ and $ Q_2\bar q_2 $, while the dashed lines show the rms radii between them. The rms radii of the ground state $ Q\bar q $ mesons with spin 0 and 1 (denoted as $ M $ and $ M^* $) are shown by the dotted lines for comparison. The vertical dotted lines indicate the physical masses of charm and bottom quarks. The red star point corresponds to the relative distance between the clusters $ Q_1\bar q_1 $ and $ Q_2\bar q_2 $ when the binding energy is adjusted to match that of the experimental $ T_{cc}(3875)^+ $.} 
		\label{fig:rvsm}
	\end{figure*}

	The  $ I(J^P)=0(1^+) $ $ QQ\bar q\bar q $ systems, in which both loosely bound molecular states and deeply bound compact states are observed in our calculations, have attracted great attention. We shall investigate very carefully how the properties of the bound states change with the heavy quark mass $ m_Q $. The binding energies of the ground state and the first excited state with varying $ m_Q $ are shown in Fig.~\ref{fig:evsm}. We see that the ground state is bound for $ m_Q $ larger than around $ 1200 $ MeV, and the binding energy increases as $ m_Q $ increases. When $ m_Q $ reaches around $ 4600 $ MeV, the first excited state also becomes a bound state. The rms radii of the two bound states with varying $ m_Q $ are shown in Fig.~\ref{fig:rvsm}. As $ m_Q $ increases, $ r^{\rm rms}_{Q_1\bar q_1} $ and $ r^{\rm rms}_{Q_2\bar q_2} $ remain stable and match the sizes of the ground state $ Q\bar q $ mesons with spin 0 and 1 (denoted as $ M $ and $ M^* $) respectively, while the other four rms radii decrease significantly. For the ground state, the rms radii between the clusters $ Q_1\bar q_1 $ and $ Q_2\bar q_2 $ are larger than $ 1 $ fm when $ m_Q $ is less than around $ 2 $ GeV, indicating a $ M^*M $ molecular configuration. In this scenario, the ground state is a loosely bound state with a binding energy $ \left|\Delta E \right|< 20$ MeV. For a larger $ m_Q $, the two heavy quarks form a compact diquark, and the ground state transforms into a deeply bound compact diquark-centered tetraquark state. For the first excited state, the rms radii between the clusters $ Q_1\bar q_1 $ and $ Q_2\bar q_2 $ decrease more slowly compared to the ground state. The first excited state remains a molecular state for $ m_Q < 6 $ GeV.   
	
	The physical interpretation for the above findings is straightforward. As $ m_Q $ increases, the kinetic energies of the heavy quarks are suppressed. The distance between two heavy quarks gets smaller and the attractive color electric interaction gets stronger, leading to a deeper bound state and a potential second bound state. Interestingly, a similar conclusion on the existence of two types of tetraquark bound states in the limit of large heavy quark mass was reached in Ref.~\cite{Allaman:2024vwn} using the Born-Oppenheimer approximation within the framework of large $ N $ QCD. Future experimental explorations of the two $ T_{bb,0(1)} $ bound states may help test theoretical predictions.
	
	\section{Summary}\label{sec:summary}
	In summary, we calculate the energy spectra of the S-wave doubly heavy tetraquark systems $ QQ^{(\prime)}\bar q\bar q$, $QQ^{(\prime)}\bar s\bar q$, and $ QQ^{(\prime)}\bar s\bar s\,(Q^{(\prime)}=b,c) $ using the AL1 quark potential model.  We apply the complex scaling method to study possible bound states and resonant states simultaneously, and the Gaussian expansion method to solve the four-body Schrödinger equation. We focus on the low-lying states below the lowest $ M(1S)M'(2S) $ dimeson threshold. The uncertainties of these tetraquark states are expected to be of the same order as those of the $ 1S $ mesons, which are around tens of MeV. 	
	
	We obtain bound states in the $ cc\bar q\bar q$, $bb\bar q\bar q$, $bc\bar q\bar q $, and $ bb\bar s\bar q $ systems. The shallow bound state $ T_{cc,0(1)}(3864) $ serves as a candidate for the experimental $ T_{cc}(3875)^+ $ state. The bound state $ T_{bc,0(2)}(7363) $ can decay strongly to $ \bar B^*D\pi $. The bound states $ T_{bb,0(1)}(10642) $ and $ T_{bc,0(1)}(7185) $ can decay radiatively. The bound states $ T_{bb,0(1)}(10491) $, $ T_{bc,0(0)}(7129) $, and $ T_{bb\bar s,1}(10647) $ can only decay weakly. In addition, a series of doubly heavy resonant states are found. We urge future experimental explorations of these predicted states.
	
    We use the rms radii to distinguish between meson molecular states and compact tetraquark states. The compact tetraquark states are further classified into three different configurations: compact even tetraquark, compact diquark-antidiquark tetraquark and compact diquark-centered tetraquark. The shallow bound states $ T_{cc,0(1)}(3864)$, $T_{bb,0(1)}(10642) $, and $ T_{bc,0(2)}(7363) $ have molecular configurations, which are QCD molecules. 
    The deeply bound states $ T_{bb,0(1)}(10491)$ and $ T_{bb\bar s,1}(10647) $ are compact diquark-centered tetraquarks, which are coined as the ``QCD helium atom" in Ref.~\cite{Liu:2019zoy}. The $ T_{bc,0(0)}(7129) $ and  $T_{bc,0(1)}(7185) $ are compact even tetraquarks, which are ideal candidates of the ``QCD hydrogen molecule" as noted in Ref.~\cite{Liu:2019zoy}. 
 
 Most of the resonant states are compact tetraquark states, except that $ T_{bb,0(1)}(10700) $ is a $ \bar B^*\bar B^* $ molecular state. The resonant states $T_{bb,0(1)}(11025)$ and $T_{bb\bar s,1}(10766)$ are considered as the radial excitations in the light degree of freedom of the bound states $ T_{bb,0(1)}(10491)$ and $ T_{bb\bar s,1}(10647) $, respectively. It is worth noting that all of the compact diquark-centered tetraquarks and compact diquark-antidiquark tetraquarks identified in our calculations are dominated by the $ \chi_{\bar 3_c\otimes3_c} $ color configuration, except for the broad resonant state $ T_{bc,0(0)}(7301) $. In these states, the attractive color electric interactions between two heavy quarks play an important role. On the other hand, mixing effect between $ \chi_{\bar3_c\otimes3_c} $ and $ \chi_{6_c\otimes\bar 6_c} $ configurations is important in compact even tetraquarks and meson molecules. Similar classifications of tetraquark configurations were made in Ref.~\cite{Allaman:2024vwn}. The compact diquark-centered tetraquarks and compact diquark-antidiquark tetraquarks resemble the type-\uppercase\expandafter{\romannumeral1} tetraquarks in Ref.~\cite{Allaman:2024vwn}, and the compact even tetraquarks and meson molecules resemble the type-\uppercase\expandafter{\romannumeral2} tetraquarks. The classifications of tetraquarks based on their color-spatial configurations help unravel the rich internal structures and various forming mechanisms of tetraquark states. 
	
	We also explore the heavy quark mass dependence of the $ T_{QQ,0(1)} $ bound states. As the heavy quark mass increases from $ 1.2 $ GeV to $ 6 $ GeV, the ground state transforms from a loosely bound molecular state to a deeply bound compact diquark-centered tetraquark state, with the emergence of a second loosely bound molecular state.  The future experimental explorations of the two $ T_{bb,0(1)} $ bound states may help test theoretical predictions and deepen our understanding of quantum chromodynamics.

	\section*{ACKNOWLEDGMENTS}
	
	We thank Zi-Yang Lin for the helpful discussions. This project was supported by the National
	Natural Science Foundation of China (No. 12475137). This
	project was also funded by the Deutsche Forschungsgemeinschaft (DFG,
	German Research Foundation, Project ID 196253076-TRR 110). The computational resources were supported by High-performance Computing Platform of Peking University.
	
	\appendix
	\section{Two definitions of root-mean-square radius }\label{app:rms}
	In our calculations, we use the decomposed nonantisymmetric wave function to calculate the rms radii. It seems more reasonable to calculate the rms radii of compact tetraquark states using the complete wave function, which considers contributions from both direct terms and exchange terms. However, our primary interest lies in the general clustering behavior of tetraquark states, rather than in specific numerical results of the rms radii, which are not experimentally observables at present. The rms radii calculated using the decomposed nonantisymmetric wave function are already capable of distinguishing between different tetraquark configurations.  To illustrate this, we compare the results of rms radii calculated using the complete wave function $\Psi $ and the decomposed non-antisymmetric term $\Psi^J_{\rm nA} $ in Table~\ref{tab:rms}. We take four states with different configurations as examples. For the compact tetraquark states $T_{bc,0(1)}(7185), T_{cc,0(1)}(4466),$ and $ T_{bb,0(1)}(10491)$, the results from $\Psi $ and $\Psi_{\rm nA} $ are similar. We can draw the same conclusion on their spatial structures from both results. However, for the molecular state $T_{bb,0(1)}(10642)$, the results from $\Psi_{\rm nA}$ can clearly demonstrate the clustering behavior of a molecular state, while the results from  $\Psi$ are more ambiguous due to the antisymmetrization. 
	\begin{table*}[htp]
		\centering
		\caption{The rms radii (in fm) calculated using the complete wave functions $\Psi$ and the decomposed nonantisymmetric wave functions $\Psi_{\rm nA} $ in Eqs.~\eqref{eq:wf_decompose1} and \eqref{eq:wf_decompose2}. The last column shows the spatial configurations of the states, where C.E., C.DA., C.DC. and M. represent  compact even tetraquark, compact diquark-antidiquark tetraquark, compact diquark-centered tetraquark and molecular configurations, respectively. }
		\label{tab:rms}
		\begin{tabular}{lcccccccc}
			\hline\hline
			State &Wave function& $ r_{Q_1\bar{q}_1}^{\mathrm{rms}} $&$ r_{Q_2\bar{q}_2}^{\mathrm{rms}} $&$ r_{Q_1\bar{q}_2}^{\mathrm{rms}} $&$ r_{Q_2\bar{q}_1}^{\mathrm{rms}} $&$ r_{Q_1Q_2}^{\mathrm{rms}} $&$ r_{\bar{q}_1\bar{q}_2}^{\mathrm{rms}} $& Configurations\\
			\hline
			$ T_{bc,0(1)}(7185) $&$\Psi_{\rm nA} $& $0.67$ & $0.66$ & $0.88$ & $0.93$ & $0.71$ & $1.00$&C.E.\\
			&$\Psi $&$0.80$&$0.82$&$0.80$&$0.82$&$ 0.75 $&$ 1.04 $&C.E.\\
			$ T_{cc,0(1)}(4466) $&$ \Psi_{\rm nA} $& $1.12$ & $1.09$ & $1.12$ & $1.13$ & $0.63$ & $0.86$&C.DA.\\
			&$ \Psi $&$ 1.10 $&$ 1.10 $&$ 1.10 $&$ 1.10 $&$ 0.63 $&$ 0.87 $&C.DA.\\
			$ T_{bb,0(1)}(10491) $&  $\Psi_{\rm nA} $&$0.68$&$0.67$&$0.70$&$0.71$&$ 0.33 $&$ 0.78 $&C.DC.\\
			&$\Psi $&$0.69$&$0.69$&$0.69$&$0.69$&$ 0.34 $&$ 0.79 $&C.DC.\\
			$ T_{bb,0(1)}(10642) $&$\Psi_{\rm nA}$& $0.66$ & $0.63$ & $2.06$ & $2.07$ & $1.98$ & $2.15$&M. $ \,(\bar B^*\bar B) $\\
			&$\Psi $&$1.52$&$1.52$&$1.52$&$1.52$&$ 1.96 $&$ 2.13 $&\\

			\hline\hline
			
		\end{tabular}
	\end{table*}
	
	In conclusion, the novel definition of rms radii, which are calculated using only the decomposed nonantisymmetric wave function, can reflect the internal spatial structure of tetraquark states more transparently. 
	
	\newpage
	\bibliography{QQqqRef}
%

\end{document}